\documentclass[journal]{vgtc}                     

\usepackage{svg}

\onlineid{1556}

\vgtccategory{Research}

\newcommand{\tool}{\textit{CellPrism}\xspace}

\title{\tool: A Visual Analytics System for Exploring AI-Driven Virtual Cells in Drug Discovery}

\author{%
  Chuhan Shi, Zijian Guo, Zelin Zang, Chengbo Zheng, Ding Ding, and Rui Sheng$^\dagger$
}

\authorfooter{
  \item
  C. Shi, Z. Guo, and D. Ding are with Southeast University. E-mail: \{chuhanshi, zijianguo, dingding-1\}@seu.edu.cn.
  \item
  Z. Zang is with Centre for Artificial Intelligence and Robotics (CAIR), HKISI-CAS and Westlake University. E-mail: zangzelin@westlake.edu.cn. 
  \item 
  C. Zheng is with the University of Queensland. E-mail: cb.zheng@connect.ust.hk.
  \item 
  R. Sheng is with the Hong Kong University of Science and Technology. E-mail: rshengac@connect.ust.hk.
  \item $^\dagger$ Corresponding author.
}

\abstract{Gene perturbation analysis plays a critical role in drug discovery by enabling researchers to investigate how interventions on specific genes influence global gene expression patterns within cells.
Recent advances in artificial intelligence (AI)-driven virtual cell models have made it possible to predict gene expression outcomes for a wide range of perturbation strategies in silico, substantially reducing reliance on costly and time-consuming biological experiments.
However, effectively exploring and interpreting the high-dimensional perturbation spaces produced by these models remains challenging due to the combinatorial nature of perturbations and the complex, cell-specific gene expression responses they generate.
In this work, we present \tool, a visual analytics system designed to support the systematic exploration of gene perturbation strategies for drug discovery.
Specifically, \tool integrates clustering-based overviews to summarize perturbation outcomes, a glyph-based representation to compactly encode gene expression patterns across cell types, and coordinated views that enable fine-grained comparison and interpretation of perturbation effects.
We demonstrate the effectiveness of \tool through a real-world case study and expert interviews.
This work highlights the value of visual analytics in bridging virtual cell modeling with expert-driven decision-making in drug discovery.
}


\teaser{%
  \centering

  \includegraphics[width=\linewidth]{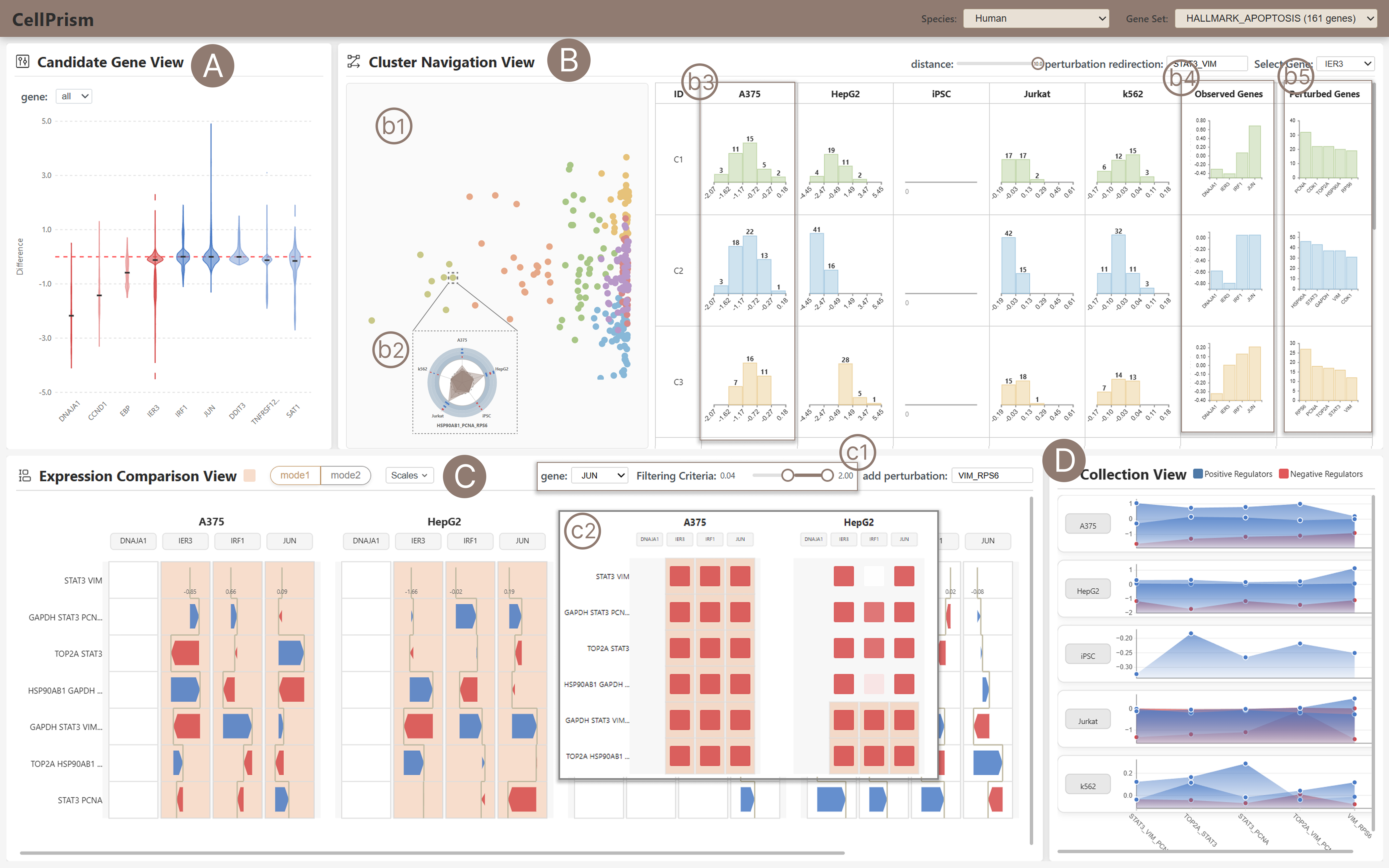}

  \par\vspace{4pt}

  \refstepcounter{figure}
  \label{fig:teaser}

  \begingroup
  \scriptsize
  \sffamily
  \raggedright
  \noindent
  Fig.~\thefigure.\enspace
  \textit{CellPrism} is a visual analytics system designed to
  facilitate cell type-specific drug target discovery and side effect
  assessment, enabling systematic exploration and screening of promising
  gene perturbation strategies. The system includes four visualization
  interfaces that guide users through a progressive, knowledge-driven
  workflow. (A) The Candidate Gene View allows experts to quickly screen
  and filter observed genes based on their statistical distribution.
  (B) The Cluster Navigation View supports exploration of the large
  perturbation space. (C) The Expression Comparison View enables detailed
  comparison and analysis of differences among perturbation strategies.
  (D) The Collection View preserves promising perturbation strategies
  selected during exploration, allowing experts to systematically compare
  candidate strategies and ultimately determine the optimal perturbation
  strategy.
  \par
  \endgroup
}

\usepackage{xspace}
\usepackage{multirow}
\usepackage{xcolor}
\usepackage{amsmath}

\newcommand{\revise}[1]{\textcolor{black}{#1}}

\definecolor{review}{HTML}{f72585}
\usepackage{marginnote}
\usepackage{adjustbox}
\global\marginparsep=9pt
\newlength{\sidecommentwidth}
\newcommand{\sidecommentline}[1]{\strut\ignorespaces#1\unskip\par}
\newcommand{\sidecommentbox}[1]{%
  \begingroup
    \setlength{\fboxsep}{2pt}%
    \setlength{\fboxrule}{0.4pt}%
    \color{review}%
    \fbox{%
      \parbox[t]{\dimexpr\sidecommentwidth-2\fboxsep-2\fboxrule\relax}{%
        \raggedright\scriptsize\sloppy
        \forcsvlist{\sidecommentline}{#1}%
      }%
    }%
  \endgroup
}
\newcommand{\sidecomment}[1]{%
  \ifdefined\revise
  \marginnote[\hspace{4pt}\sidecommentbox{#1}]{\hspace{-4pt}\sidecommentbox{#1}}%
  \fi
}

\renewcommand{\sidecomment}[1]{}

\graphicspath{{figs/}{figures/}{pictures/}{images/}{./}} 

\usepackage{tabu}                      
\usepackage{booktabs}                  
\usepackage{lipsum}                    
\usepackage{mwe}                       
\usepackage{ccicons}  

\usepackage{mathptmx}                  

\AtBeginDocument{%
  \crefname{figure}{Fig.}{Figs.}%
  \Crefname{figure}{Fig.}{Figs.}%
}

\begin{document}



\maketitle

\section{Introduction}

Drug discovery for specific cell types and the evaluation of potential side effects typically require extensive biological experiments, which are costly in terms of time, finances, and human resources~\cite{khikhmetova2025advances, dimasi2016innovation, schuhmacher2023analysis}. 
One crucial aspect of this process is identifying \textbf{perturbed genes} (i.e., genes that are deliberately manipulated through interventions) to regulate the expression of downstream \textbf{observed genes} that are directly associated with the cellular phenotypes of interest, while minimizing unintended effects on other cellular functions~\cite{kalter2025off}.
\sidecomment{R2C6}
\revise{However, experimentally testing all possible perturbed genes is time-consuming since researchers can perturb individual or combined genes (e.g., perturbing STAT3 alone, perturbing PCNA alone, or perturbing STAT3 and PCNA together), generating hundreds or even thousands of candidate perturbation strategies.}
Currently, virtual cell models, especially single-cell perturbation models, offer a promising alternative~\cite{adduri2025predicting, yu2025perturbnet, tang2025cellforge, klein2025cellflow, theodoris2023transfer}.
Those models are trained on large-scale gene perturbation data to predict the resulting expression of observed genes for given perturbation strategies.
This enables researchers to rapidly evaluate potential interventions without relying solely on costly biological experiments.

\sidecomment{R2C4}
\revise{Despite the capability of virtual cell models to simulate specific perturbation outcomes, fully automated optimization among the candidate strategies is impractical. Although experts have clear high-level therapeutic goals (e.g., effectively killing tumor cells), there is rarely a ``perfect'' strategy; instead, each candidate often comes with its own set of biological risks. Experts must carefully balance the trade-offs between therapeutic efficacy and potential side effects to identify the most acceptable outcome. Consequently, rather than relying on predefined thresholds or one-size-fits-all criteria, experts often need to heuristically compare alternative perturbation strategies and determine which changes are biologically acceptable in a real-world scenario based on the predicted effects.
However, this comparison and exploration process will introduce immense cognitive load, as the simulated effects are multi-dimensional and complex. Each candidate perturbation strategy will trigger expression changes across multiple genes that vary substantially across different cell types~\cite{lotfollahi2019scgen}, making it difficult for experts to track and understand these variations manually.
}

Various visualization tools help researchers explore and analyze gene expression data \cite{simon2017visexpress, patil2023scviewer, wei2025perturbase, zhou2019networkanalyst, wolf2018scanpy}.
For example, VisExpress~\cite{simon2017visexpress} utilizes gene fingerprint visualization to achieve interactive visual exploration for pairwise combination comparison and quality perception of large-scale differentially expressed gene data. Additionally, scViewer~\cite{patil2023scviewer} employs a multi-view collaborative approach to analyze gene expression data. Vis-SPLIT~\cite{roper2023vis} can help analyze gene expression data, leveraging PCA projections and interactive multi-view feedback. TrajLens~\cite{wang2026TrajLens} has designed a fusion view of cell distribution and development directions across multiple samples, helping biologists explore the spatial evolution patterns of cell populations on trajectories in cross-sample single-cell RNA sequencing data.
However, they remain limited in supporting researchers’ exploration of the complex relationships between perturbations and observed gene expression profiles.

To address this challenge, we introduce \tool, a visual analytics system designed to support exploratory reasoning over large-scale gene perturbation spaces for drug discovery. Rather than focusing on isolated perturbation outcomes, \tool enables experts to iteratively narrow down candidate perturbations by jointly considering their effects on target cells for apoptosis induction 
\sidecomment{R2C1}
\revise{ (i.e., promoting the elimination of harmful cells) }
and protected normal cells.
Specifically, our system supports a progressive exploration process that guides experts from a global overview of the perturbation space to detailed analysis. Additionally, we design a radar-based glyph that compactly displays the expression levels of observed genes across different cells for each perturbation, while explicitly encoding the contributions of individual perturbed genes to these expression changes, enabling experts to reason about how different perturbation components jointly drive cell-type–specific responses.
Overall, our contribution can be summarized as follows:

\begin{itemize}[leftmargin=*, topsep=4pt, itemsep=2pt]
\item We derived five design requirements through interviews with three experienced domain experts to support gene perturbation exploration in drug discovery.
\item We developed a visual analytics system, \tool, to assist domain experts in gene perturbation exploration for drug discovery.
\item We validated the effectiveness of our system based on a real-world dataset through a case study and an expert interview.
\end{itemize}

\section{\revise{Background}}
\sidecomment{R2C5}
In this section, we introduce three core concepts and their relationships: \textit{\textbf{perturbed and observed genes}} define the inputs and outputs of gene interventions, \textit{\textbf{gene expression differences}} capture how observed genes respond to these interventions relative to baseline states, and \textit{\textbf{single-cell perturbation models}} provide predictive estimates of these expression differences across multiple cell types. Together, these concepts form the foundation for exploratory perturbation analysis.

$\diamond$ \textit{\textbf{Perturbed and observed genes.}} 
Within cell-based drug discovery, gene perturbation analysis distinguishes between two roles of genes: those that are directly intervened on and those whose expression changes are monitored. \emph{Perturbed genes} refer to genes that are deliberately manipulated through interventions, serving as controllable inputs to the system. A single perturbation may involve one gene or a combination of multiple genes.
In contrast, \emph{observed genes} are not directly manipulated but are selected to characterize cellular responses to perturbations. Their expression patterns provide gene-level evidence of how cells react to different interventions. Collectively, the expression of multiple observed genes is used to infer higher-level cellular phenotypes, such as survival or apoptosis.

$\diamond$ \textit{\textbf{Differential gene expression.}} 
Differential gene expression measures changes relative to a baseline state. For each perturbation and cell type, the system computes gene expression differences that quantify how the expression levels of observed genes deviate from their baseline values. These differences provide a compact representation of how a perturbation shifts a cell’s transcriptional state toward or away from desired therapeutic outcomes, such as cancer cell death or the preservation of normal cells.
Importantly, the interpretation of expression differences depends on both the cell type and the functional role of each gene. For example, increased expression of pro-death genes or decreased expression of pro-survival genes may indicate effective induction of apoptosis in cancer cells, whereas similar changes in normal cells may raise safety concerns. As a result, experts must reason about the direction and magnitude of expression differences in a cell-type–specific and gene-aware manner when evaluating perturbation effects.

$\diamond$ \textit{\textbf{Single-cell perturbation model.}} 
We conceptualize a single-cell perturbation model as a data-driven mapping that captures how gene perturbations reshape cellular gene expression profiles at the single-cell level. 
Specifically, a cell is represented as a high-dimensional vector of gene expression values, and a perturbation corresponds to an intervention applied to one or more genes. 
Formally, single-cell perturbation data are represented as a gene expression variation matrix
$\mathbf{X} \in {R}^{N \times M}$,
where each row corresponds to an individual cell and each column corresponds to a gene.
Rather than using absolute expression values, we focus on differential gene expression, defined as the difference between post-perturbation expression and a pre-perturbation baseline.
This representation aligns with common analytical practice in biology, where the effect of a perturbation is assessed through relative changes rather than raw expression levels.

\begin{figure}[ht]
    \centering
    \includegraphics[width=\linewidth]{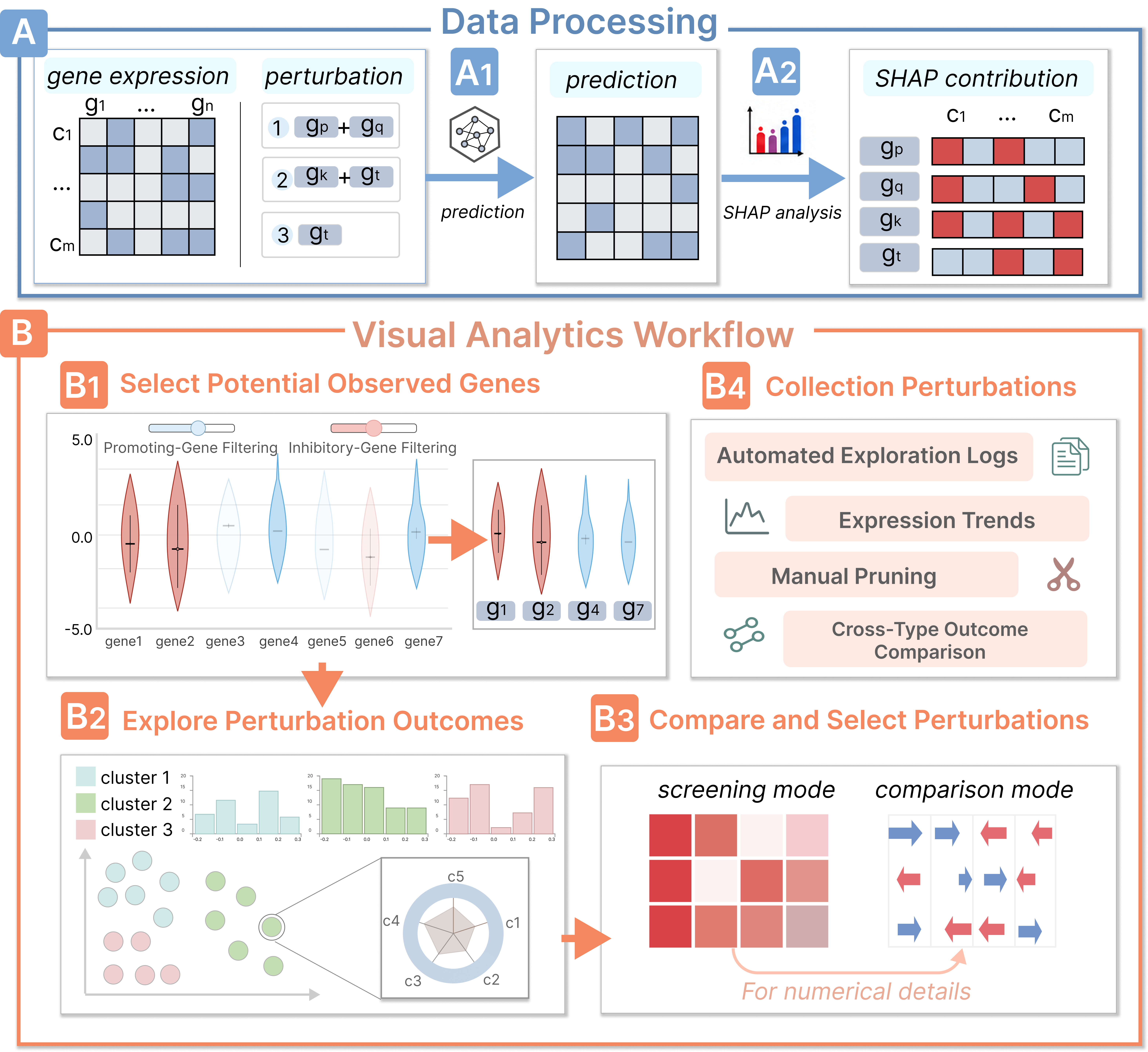}
    \caption{The integrated framework of \tool, encompassing (A) the data processing pipeline and (B) the visual analytics workflow. We first leverage virtual cell models to process the input gene expression matrix and perturbation labels, (A1) predicting gene differential expression and (A2) calculating the contribution of each perturbed gene in different cells. Subsequently, the visual analytics workflow guides users through four progressive exploration stages: (B1) filtering and selecting observed genes; (B2) obtaining an overview of perturbation outcomes; (B3) comparing promising perturbation strategies in detail; and (B4) evaluating selected perturbation strategies for the final decision.}
    \label{fig:main}
\end{figure}

A perturbation is encoded as a set of perturbed genes applied to each cell.
Given an input consisting of (1) a baseline gene expression matrix and (2) perturbed genes, the goal of a single-cell perturbation model is to predict the resulting differential gene expression matrix (\cref{fig:main}-A1).
In other words, the model learns a transformation:
\[
(\mathbf{X}_{\text{baseline}}, \mathbf{p}) \;\rightarrow\; \Delta \mathbf{X},
\]
where $\mathbf{p}$ denotes one or more perturbed genes, and $\Delta \mathbf{X}$ denotes the predicted expression changes across all genes and cells.

\section{Related Works}
In this section, we examine prior works related to our research and elaborate on the limitations of existing approaches.

\subsection{Human-AI Collaboration for Drug Discovery}

Drug discovery is a multi-stage process that ranges from drug target identification to candidate drug validation \cite{emmerich2021improving, hughes2011principles, mohs2017drug, sheng2025Trialcompass}. The essence of drug therapy is that drug molecules treat diseases by binding to specific targets \cite{pushpakom2019drug}. If the protein encoded by a certain gene is identified as the target of the drug, then the drug discovery and screening can logically be transformed into the exploration and discovery of the effect of specific gene perturbation. Traditional methods for drug target identification rely on experimental screening; however, it is confronted with disadvantages such as high costs and long cycles \cite{khikhmetova2025advances, morgan2011cost}. In recent years, the rise of AI has provided a new method for target identification. 
For example, transformer approaches like STATE have been used to predict perturbation effect \cite{adduri2025predicting}. Cellforge \cite{tang2025cellforge} utilizes a multi-agent framework to customize generative models based on research objectives and datasets. Rbio1 \cite{istrate2025rbio1} trains the LLM via reinforcement learning, using prediction results from other biological models as rewards. 
However, given the current limitations in AI accuracy, it is unrealistic to rely on fully automated systems for this task. As a result, human–AI collaboration has become increasingly important in drug discovery. Some systems assist researchers by mining large-scale biomedical literature, helping uncover potential associations among genes, drugs, and diseases to support hypothesis generation \cite{wong2021search, wilson2018automated, cheerkoot2021literature, frijters2010literature}.
Other approaches operate at the molecular level, where AI models are used to predict properties such as activity, toxicity, or binding affinity, enabling early-stage candidate screening \cite{li2022deep, wu2018moleculenet, mayr2016deeptox, jimenez2018k}.
More recent work has begun to explore human–AI collaboration at the cellular level, incorporating cell-based data to reason about drug effects \cite{liu2024vibrant}.
However, existing approaches rarely consider how virtual cell models can be integrated into interactive workflows that support joint exploration and decision-making between humans and AI. This gap limits researchers’ ability to systematically examine perturbation effects within a cell-level context. Our work addresses this limitation by leveraging virtual cells as a central medium for human–AI collaboration in drug discovery.


\subsection{AI Algorithms for Virtual Cells}

Virtual cells have emerged as a powerful computational model to simulate biological functions and interactions within a cell \cite{bunne2024build}. Existing virtual cell models can be broadly categorized into two strategies: knowledge-incorporated models and data-driven models. Knowledge-incorporated models embed prior biological knowledge, such as curated pathways or graph structures, into the modeling process to constrain predictions \cite{fang2025cell, roohani2024predicting, ma2025ai, elmarakeby2021biologically}. For example, Cell-Graph Compass incorporates graph-based pretraining to encode biological structure as prior knowledge \cite{fang2025cell}. The GEARS model incorporates a gene-gene knowledge graph as prior knowledge to simulate and predict results through deep learning \cite{roohani2024predicting}.
In contrast, data-driven models rely on large-scale single-cell gene perturbation datasets to learn cellular responses directly from data \cite{yu2025perturbnet, adduri2025predicting, lotfollahi2019scgen, cui2024scgpt}. Models such as PerturbNet and STATE are trained on extensive perturbation experiments and can predict gene expression profiles given expert-specified perturbations \cite{yu2025perturbnet, adduri2025predicting}. Despite differences in modeling approaches, these methods primarily output predicted high-dimensional gene expression profiles that are difficult for experts to interpret and leverage for deeper reasoning or complex decision-making in drug discovery.
To better support expert analysis, it is therefore necessary to complement advanced virtual cell models with visualization techniques that make predicted perturbation effects more interpretable and actionable within the drug discovery process.

\subsection{Visual Analytics for Gene Expression Data}

Gene expression data are typically characterized by large volume and high dimensionality. Some researchers develop visualization analysis tools to visualize high-dimensional data and assist experts in understanding it~\cite{Sheng2026CellScout, kleverov2024phantasus, megill2021cellxgene, fernandez2017clustergrammer, wang2026TrajLens}. For example, Phantasus \cite{kleverov2024phantasus} is a web application that supports the complete workflow of gene expression analysis, from data preprocessing and exploration to differential gene expression and pathway analysis. Cellxgene \cite{megill2021cellxgene} links embedding plots and gene feature histograms to alleviate information overload. Clustergrammer \cite{fernandez2017clustergrammer} enables interactive visual analysis of static heatmaps for data uploaded by users. Other researchers have developed tools dedicated to specific biological tasks, such as single-cell trajectory inference \cite{chen2019single, smolander2023cell} and alternative splicing analysis \cite{wong2024splicewiz}. StratomeX enables the exploration of candidate cancer subtypes through the visualization of multi-omics data \cite{lex2012stratomex}. 
Such analysis methods are typically limited to known data from real experiments. These methods depend on dimensionality reduction for projection into 2D space, which is inadequate for analyzing many-to-many relationships. This limitation becomes particularly acute in virtual cell simulations, where a single genetic perturbation or multi-gene combinations can trigger cascading effects across thousands of downstream genes under various cell states, yielding a complex, non-linear relationship. However, there is currently no visualization tool that can address the need to fully display the complex relationships among virtual-cell-generated data and support experts in exploring them from multiple perspectives. To address these problems, we have developed a visual analysis tool, \tool.

\revise{\section{Design Study}}
%
Our goal in designing a visual analytics system is to support domain experts in exploring and reasoning about high-dimensional perturbation outcomes generated by virtual cell models for drug discovery. Researchers need to go beyond identifying perturbations that produce desirable expression changes and further assess how these effects propagate across multiple genes and cell types, whether the observed patterns align with their biological hypotheses, and how alternative perturbation strategies compare under different analytical perspectives. Such reasoning is particularly challenging due to the combinatorial nature of gene perturbations and the complex, many-to-many relationships between perturbations, genes, and cellular contexts.

To ground the design of our system, we conducted a formative interview with three domain experts (E1–E3). E1 and E2 are Ph.D. students with four years of experience in computational biology and single-cell analysis, both actively working on gene perturbation modeling and downstream biological interpretation. E3 is a postdoctoral researcher with eight years of experience in proteomics and functional genomics, with extensive involvement in analyzing large-scale omics data for mechanism discovery.
\sidecomment{R2C8}
\revise{Based on experts' feedback, we derived a preliminary set of design requirements that informed our initial design of the system. Then we met with these three experts biweekly for six months and invited them to co-design our system iteratively. Through this co-design process, we deepened our understanding of experts' analytical practices and incorporated their feedback into the system implementation. Finally, we consolidated the feedback into five design requirements, which include both the initial design requirements derived from the formative interview and the new ones subsequently identified through the iterative design process.}
\revise{Our study was approved by the School of Computer Science and Engineering, Southeast University, and informed consent was obtained from each expert.}
\begin{itemize}[leftmargin=2.5em]
    \item[\textbf{DR1}] \textbf{Support the identification of relevant observed genes.} During gene perturbation analysis, a single perturbation can induce expression changes in a large number of genes. Although experts typically have high-level analytical goals, such as inducing apoptosis or preserving cell viability, the set of genes associated with these processes is often broad. For instance, more than 160 genes have been reported to be related to cell death in humans, and monitoring all of them simultaneously is neither practical nor informative.
    Instead of exhaustively examining all responsive genes, experts need to identify a subset of observed genes that are most relevant to their analytical objectives. In practice, experts often prioritize genes that exhibit pronounced expression differences or are known to play decisive roles in regulating the target cellular processes, while deemphasizing genes with marginal or noisy responses. 
    \sidecomment{R2C9}
    \revise{Therefore, the system needs to support interactive filtering based on how much the genes' expression levels change after a perturbation, helping experts quickly hide the inactive genes and focus on the ones with major changes.}

    \item[\textbf{DR2}] \textbf{Provide an overview of gene expression outcomes after experts specify a set of perturbation strategies}.
    Experts first decide on a set of genes that can be perturbed for experimentation. However, even a relatively small set of candidate perturbed genes can generate a large number of perturbation strategies through different combinations. As a result, experts often struggle to form a holistic understanding of the resulting gene expression outcomes across perturbation strategies, making it difficult to determine where to begin detailed analysis.
    As E1 noted, \textit{``Before looking into individual perturbations, I want to first know what kinds of expression patterns these perturbations generally produce, so that I can decide where to start.''}
    As E3 noted, \textit{``This requires considering different perturbations, different cell types, and even genes within individual cells, which is very hard for us to reason about simultaneously.''}
    Therefore, experts need a high-level overview that summarizes and groups perturbation outcomes based on their gene expression patterns.

\item[\textbf{DR3}] \textbf{Reveal contributions of each perturbed gene for each observed gene}. 
Some perturbation strategies involve simultaneously perturbing multiple genes, which collectively influence the resulting gene expression profiles.
In such cases, experts need to understand how much each perturbed gene contributes to the observed expression outcomes in order to identify the dominant drivers of the perturbation effects.
As E3 noted, \textit{``When multiple genes are perturbed together, it is hard to tell which one is actually responsible for the changes we observe.''}
Without explicit support for attributing effects to individual genes, experts must rely on manual inspection or prior assumptions, making it difficult to interpret model predictions, compare alternative perturbation strategies, or decide how to refine perturbation designs.
Therefore, the system should reveal gene-level contribution information for each perturbation strategy, enabling experts to disentangle the roles of individual genes.

\item[\textbf{DR4}] \textbf{Enable fine-grained comparison of differential gene expression across different perturbations}. 
In practice, experts often identify a subset of perturbations of interest and wish to examine them side by side at a finer level of detail.
As E2 explained, \textit{``In drug discovery, we need to closely compare how different perturbations affect the same cells.''}
Such analyses require experts to focus on a shared cellular context and directly compare how gene expression responses differ across perturbation strategies.
As E1 echoed this view, \textit{``Given multiple cells and diverse perturbation strategies, comparing their effects in a structured and reliable way is very challenging.''}
Therefore, the system should explicitly support fine-grained comparison of gene expression responses across multiple perturbations under a shared cellular context, enabling experts to systematically contrast perturbation effects and draw more reliable conclusions.

\sidecomment{Second Round Comments}
\item[\textbf{DR5}] \textbf{Support iterative exploration and \revise{collective review among candidate perturbation strategies}}. 
After identifying promising perturbation strategies, experts typically refine their exploration based on insights gained from earlier analysis.
This process is iterative and exploratory, rather than a linear search with a clearly defined objective.
As E1 noted, \textit{``When searching through different perturbations, we often only have a vague goal in mind. To make a final decision, we need to make trade-offs among candidate perturbations.''}
Without explicit support for \revise{retaining and comparing perturbation strategies identified during exploration}, experts may \revise{lose track of promising candidates from earlier iterations as their goals and evaluation criteria gradually evolve}, making it difficult to conduct systematic comparisons or justify final choices.
Therefore, the system should support \revise{saving and revisiting promising strategies during exploration.}
Such support enables experts to compare alternatives across iterations and collectively evaluate the trade-offs among them, ultimately facilitating more informed and confident perturbation selection.
\end{itemize}{}

\section{System Design}

We present \tool (\cref{fig:teaser}), a visual analytics system designed to facilitate the exploration and selection of optimal perturbations for cell-type-specific drug target discovery and side effects assessment. \tool integrates four views and enables a workflow fit into experts' needs. The workflow begins in the Candidate Gene View (\cref{fig:teaser}-A), where experts filter and identify observed genes of interest \textbf{(DR1)}. 
Subsequently, users explore the perturbation space (\cref{fig:teaser}-b1) within the Cluster Navigation View (\cref{fig:teaser}-B), where they can obtain an overview of the clustering distributions. Meanwhile, the radar-based glyphs (\cref{fig:teaser}-b2) enable a direct comparison and evaluation of perturbations \textbf{(DR2, DR3)}.
Users can then select the perturbations of interest and add them to the Expression Comparison View (\cref{fig:teaser}-C), enabling a detailed comparison of the perturbation data \textbf{(DR4)}. 
During the process of exploration, the Collection View (\cref{fig:teaser}-D) preserves promising perturbation strategies selected during exploration to support users in comparing and selecting the most appropriate perturbation \textbf{(DR5)}.

\subsection{Candidate Gene View}
\begin{figure}[h]
    \centering
    \includegraphics[width=\linewidth]{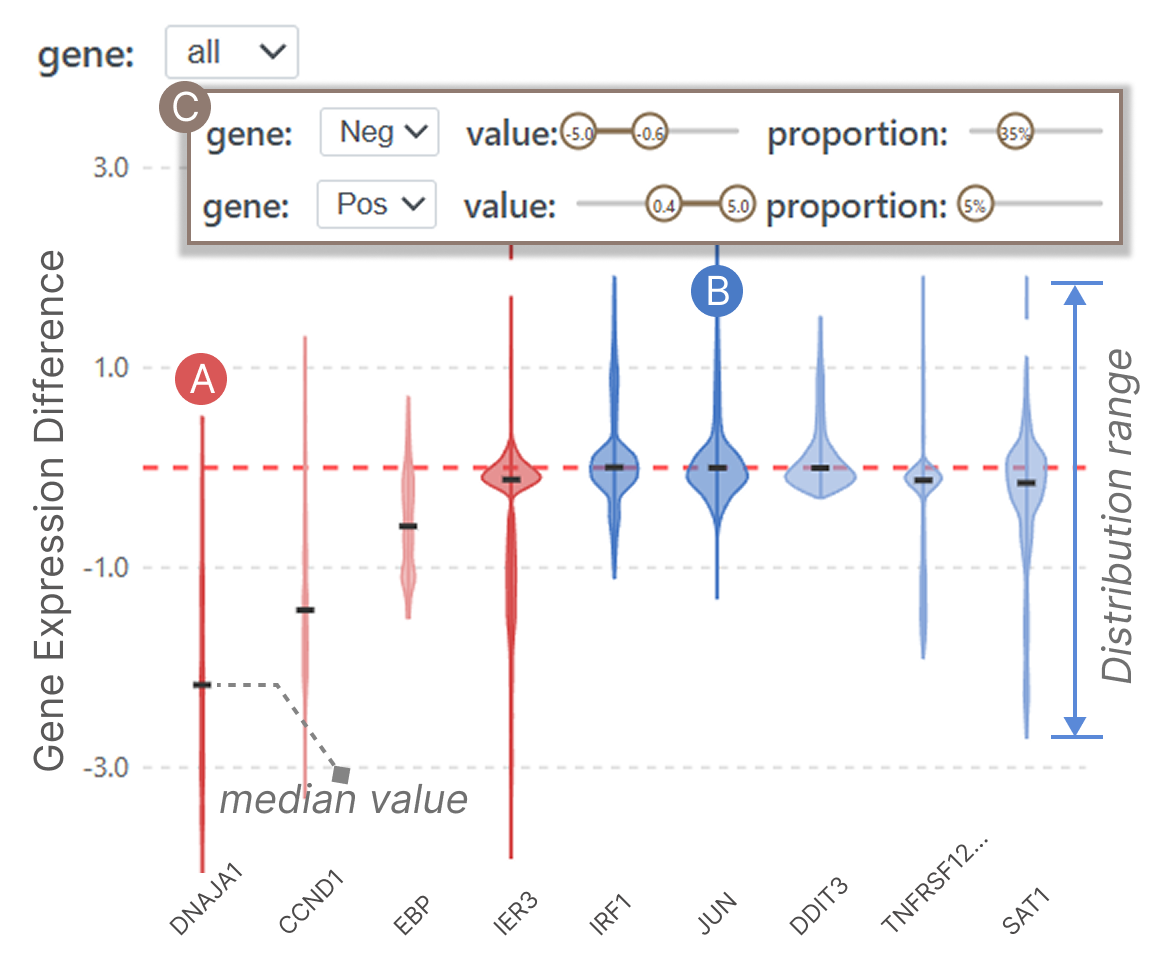}
    \caption{Distribution of differential gene expression visualized with side-by-side violin plots. Each violin represents one gene, where the y-axis encodes expression difference relative to a reference population and the width indicates data density. Red violins (A) denote genes that inhibit the cell state, while blue violins (B) denote promotive genes. \revise{(C) Range sliders separately filter inhibitory (i.e., Neg) and promotive (i.e., Pos) genes by specifying a range of gene expression values and the minimum proportion of perturbation outcomes that fall within that range.}}
    \label{fig:violin}
\end{figure}

The Candidate Gene View \revise{(\cref{fig:teaser}-A)} enables rapid screening of potential observed genes by visualizing the distributions of their differential gene expression across all user-specified perturbation strategies, as predicted by the perturbation model. By aggregating expression outcomes over the perturbation set, the view provides an overview of how each gene responds to perturbations.
\sidecomment{R2C7}
Specifically, differential gene expression is visualized using side-by-side violin plots, a representation familiar to domain experts and well-suited for revealing distributional patterns\revise{, to assist experts in filtering genes based on the overall differential expression distributions (DR1) (\cref{fig:main}-B1)}. In the violin plot (\cref{fig:violin}), each violin corresponds to an individual observed gene. The y-axis encodes the gene expression difference relative to the baseline state, while the width of the violin reflects the density of perturbation strategies that induce similar expression changes.
To support quantitative interpretation, each violin is annotated with its median expression difference, allowing experts to assess the overall direction of regulation. In addition, the vertical extent of the violin indicates the range of expression responses, helping experts identify genes with stable versus highly variable behavior across perturbations.
Genes are color-coded by their functional roles in regulating cell states: red violins  \sidecomment{R2C13} \revise{(\cref{fig:violin}-A)} represent genes that inhibit the target cell state, whereas blue violins \revise{(\cref{fig:violin}-B)} represent genes that promote it. This encoding helps experts reason more systematically about the magnitude, direction, and consistency of expression differences across candidate observed genes.

\textbf{Interaction.} Experts can interactively filter candidate genes using a range slider \sidecomment{R2C12}\revise{(\cref{fig:violin}-C)}. Because the selection criteria differ between cell state–promoting and inhibitory genes, the system provides separate filtering thresholds for each category. These thresholds are flexibly adjusted by experts based on their biological prior knowledge and real-time visual feedback from the underlying data distributions.

\subsection{Cluster Navigation View}
The Cluster Navigation View (\cref{fig:teaser}-B) provides a multi-scale overview for experts to make sense of the vast, high-dimensional perturbation space. To bridge the gap between global distribution and local inspection, we implement a two-level hierarchical exploration strategy (global-to-local), enabling experts to narrow down candidates from thousands of perturbations to a manageable subset for rigorous assessment \sidecomment{R2C7} \revise{(DR2)}.
Specifically, we combine a cluster-level overview for coarse-grained screening with a glyph-based local inspection for detailed assessment of individual perturbations (\cref{fig:main}-B2).

\begin{figure}[h]
    \centering
    \includegraphics[width=1.0\linewidth]{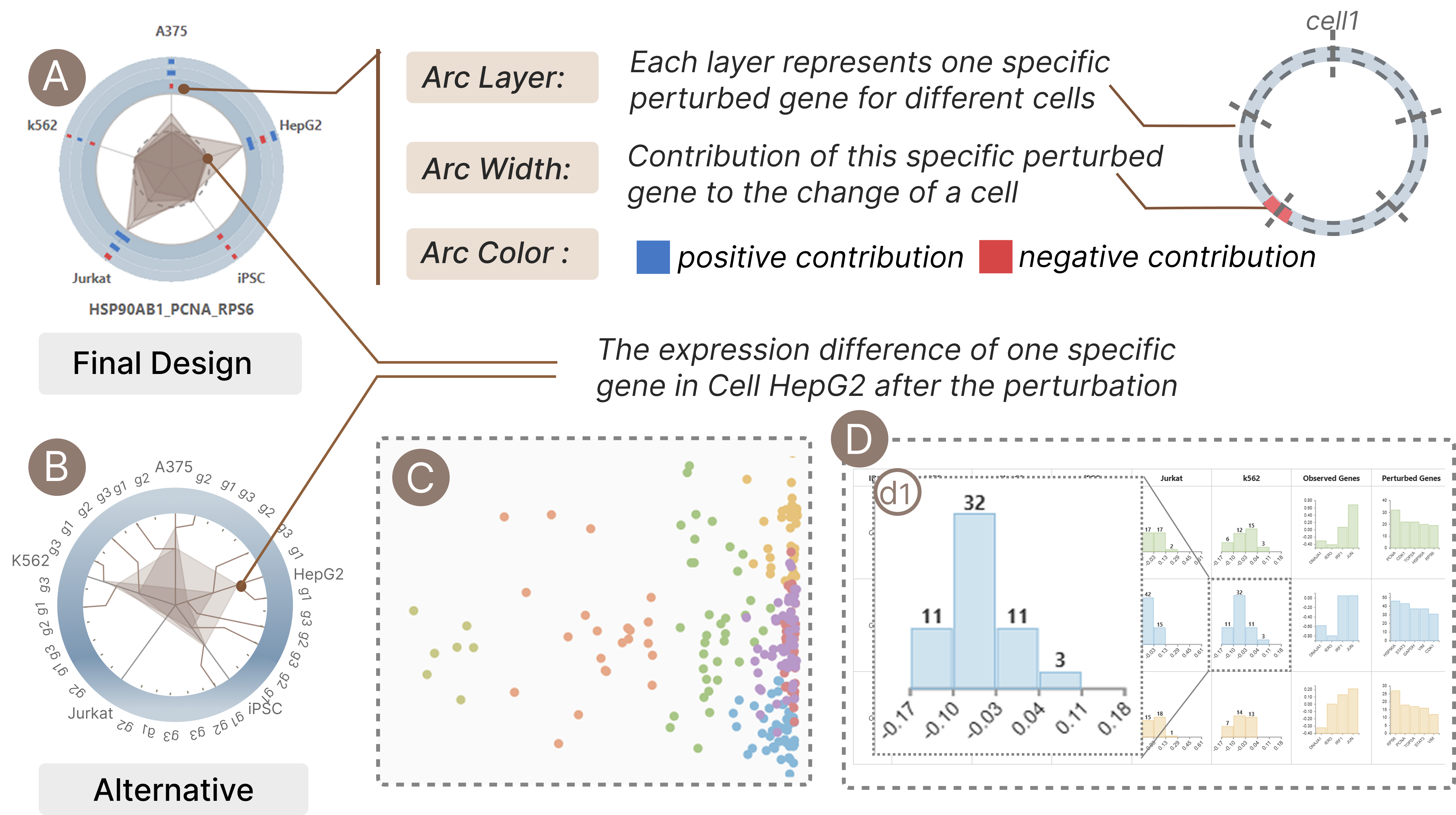}
    \caption{(A) Radar-Based Glyph: When users zoom into a region of interest, perturbations in the projection plot smoothly transition into radar ring glyphs, revealing fine-grained, cell- and gene-level features. (B) Alternative Design: An early design iteration of the radar ring glyph that was explored but ultimately discarded. (C) Cluster Projection Plot: The initial overview showing the distribution of perturbation strategies in the projected space, where each point represents an individual perturbation. (D) Cluster Summary Table: A tabular overview summarizing cluster-level statistics to support high-level comparison and selection.}
    \label{fig:cluster_projection_plot}
\end{figure}

\textbf{Cluster Overview.}  
The Cluster Overview supports coarse-grained screening of the high-dimensional perturbation space, helping experts quickly identify promising regions for further inspection. It consists of two coordinated components that provide spatial context and summary statistics.
The cluster projection plot (\cref{fig:cluster_projection_plot}-C) visualizes the outcomes of perturbations, where each point represents a perturbation strategy and spatial proximity indicates similarity in gene expression differences. This global layout allows experts to perceive the structure of the perturbation outcomes and identify regions of interest that warrant further attention.
Complementing the cluster projection plot, a tabular summary view (\cref{fig:cluster_projection_plot}-D) provides cluster-level statistics to facilitate targeted filtering, where each row corresponds to a cluster. 
\sidecomment{R2C16}
\revise{The table summarizes each cluster from three complementary perspectives: cell-type-specific expression distributions, observed gene responses, and perturbation composition. For each cell-type column, such as the A375 column (\cref{fig:teaser}-b3), each table cell contains a histogram showing how perturbation strategies in the cluster are distributed across the predicted expression-change range of the observed gene in the corresponding cell type. In each histogram (\cref{fig:cluster_projection_plot}-d1), the horizontal axis represents differential gene expression intervals, while the vertical axis represents the number of perturbation strategies falling into each interval. By checking whether the histogram is mainly distributed on the negative or positive side, experts can identify the dominant response direction of a cluster in each cell type.
The observed gene column (\cref{fig:teaser}-b4) summarizes the average expression response of observed genes within each cluster. In each bar chart, the x-axis lists the observed genes, and the y-axis encodes their average differential expression across perturbation strategies in the cluster. Through these charts, experts can rapidly exclude clusters that do not satisfy basic criteria on expression direction or magnitude. The perturbed gene column (\cref{fig:teaser}-b5) describes the perturbation composition of each cluster. In each bar chart, the x-axis lists perturbed genes, and the bar height encodes how many perturbation strategies in the cluster involve each gene. This allows experts to identify which perturbation components are common in a cluster before inspecting individual perturbations.}
Together, these coordinated views allow experts to progressively narrow down the search space while maintaining awareness of global structure and local variability.

\textbf{Radar-based Glyph.}  
To support the local inspection of individual perturbations, we employ a semantic zoom mechanism.
\sidecomment{R2C15}
\revise{As users zoom into a region of interest in the Cluster Overview, each point in the projection plot smoothly transitions into a detailed radar-based glyph (\cref{fig:cluster_projection_plot}-A).}
Each glyph abstracts complex multi-gene–multi-cell relationships into compact geometric patterns, enabling efficient assessment of perturbation effects at the individual level.
In the glyph, each axis corresponds to an individual cell, and values along the axis encode the differential gene expression of \revise{the observed genes retained after filtering in the Candidate Gene View for that cell. 
\sidecomment{R2C17}
Early design iterations used distinct colors to differentiate genes. 
However, expert feedback indicated that during exploration, users prioritize rapid, coarse-grained assessment over precise gene-by-gene inspection. Also, experts are mainly interested in the outermost contours because they indicate the largest deviations from the shared central baseline and thus highlight the most salient perturbation-induced responses. To reduce cognitive load, we therefore adopt a uniform color scheme and connect the same gene across adjacent cells to form polygonal contours.}
By observing how these contours deviate from a shared central baseline, experts can quickly gauge the overall impact of a perturbation across cells without explicit numerical comparison.
To support interpretability, we augment the radar-based glyph with a contribution explanation layer based on \revise{SHapley Additive exPlanations (SHAP) }values~\cite{dwivedi2023explainable, alicioglu2022survey, rozemberczki2022shapley}
\sidecomment{R2C1}
\revise{, which quantify how each perturbed gene contributes to the expression difference of individual cells (\cref{fig:main}-A2).}
Contribution information is displayed along the periphery of the glyph using an arc, where color encodes contribution direction (blue for positive contribution and red for negative contribution), and arc length represents contribution magnitude. 
Perturbed genes are arranged radially from the innermost to the outermost ring according to the perturbation's gene composition. The color intensity of each ring reflects the average contribution of the corresponding gene across all cells, providing a summary of how individual perturbed genes influence observed gene expression.
Through this glyph, experts can quickly identify which perturbed genes play a dominant role within a multi-gene perturbation strategy, enabling them to explore potential alternative perturbation strategies \sidecomment{R2C7} \revise{(DR3)}.

\textbf{Interaction.} 
To streamline exploration of the perturbation space, we design a set of coordinated interaction mechanisms that enable experts to screen, search for, and precisely locate target perturbations.
To reduce visual interference when navigating dense clusters, the system establishes a coordinated link between the cluster overview table and the cluster projection plot. Selecting a cluster in the table automatically deemphasizes unrelated clusters by reducing their opacity in the projection plot, allowing experts to focus on the target subspace while preserving the global distribution as contextual background, thereby mitigating visual clutter.
In addition, recognizing that experts often validate hypotheses based on prior biological knowledge or insights derived from SHAP-based contribution analysis, we provide a targeted retrieval function that allows users to directly locate specific perturbations without exhaustive manual exploration.
Finally, to maintain analytical continuity across views, the system supports cross-view coordination: selecting a radar-based glyph in the Cluster Projection Plot populates its detailed expression profile in the Expression Comparison View (\cref{fig:teaser}-C), enabling fine-grained, side-by-side comparison.

\textbf{Justification.} 
In early design iterations, we explored a leaf-shaped glyph–inspired visualization (\cref{fig:cluster_projection_plot}-B) to encode the complex relationships between cells, genes, and their contribution levels within a single perturbation. In this design, each edge corresponded to a cell, while semi-arcs on opposite sides distinguished positive and negative gene expressions. Observed genes were initially arranged in a fixed order along the cell boundaries, with polylines connecting gene labels to their corresponding expression values.
As the number of observed genes increased, these connectors became heavily entangled, resulting in substantial visual clutter. To improve visual clarity and reduce cognitive burden, we introduced a dynamic reordering strategy that sorted genes by the magnitude of expression differences, and summarized gene contributions using a single gradient ring along the outer boundary. Although these refinements improved aesthetic clarity and alleviated line tangling, the resulting layout still failed to support rapid, coarse-grained assessment of cell states. Moreover, it hindered cross-cell comparison of the same gene and lacked the resolution needed to attribute gene-level contributions to individual cells.
As a result, we abandoned this alternative design in favor of the current radar-based glyph, which better supports both high-level judgment and fine-grained comparative analysis across cells and genes.

\subsection{Expression Comparison View}
The Expression Comparison View (\cref{fig:teaser}-C) supports systematic, side-by-side comparison of candidate perturbations to help experts identify strategies that best meet their experimental goals.
Specifically, the Expression Comparison View provides two complementary modes: \emph{Screening Mode} and \emph{Comparison Mode} (\cref{fig:main}-B3).
\emph{Screening Mode} is designed for rapid filtering over multiple potential perturbation strategies selected from the Cluster Navigation View, allowing experts to quickly exclude ineffective candidates based on their overall expression patterns relative to a predefined filtering threshold. In contrast, \emph{Comparison Mode} supports fine-grained, gene-level inspection of a small set of promising perturbations, enabling precise comparison of expression differences and trajectories across perturbation strategies  \revise{(DR4)}.

\begin{figure}[h]
    \centering
    \includegraphics[width=1.0\linewidth]{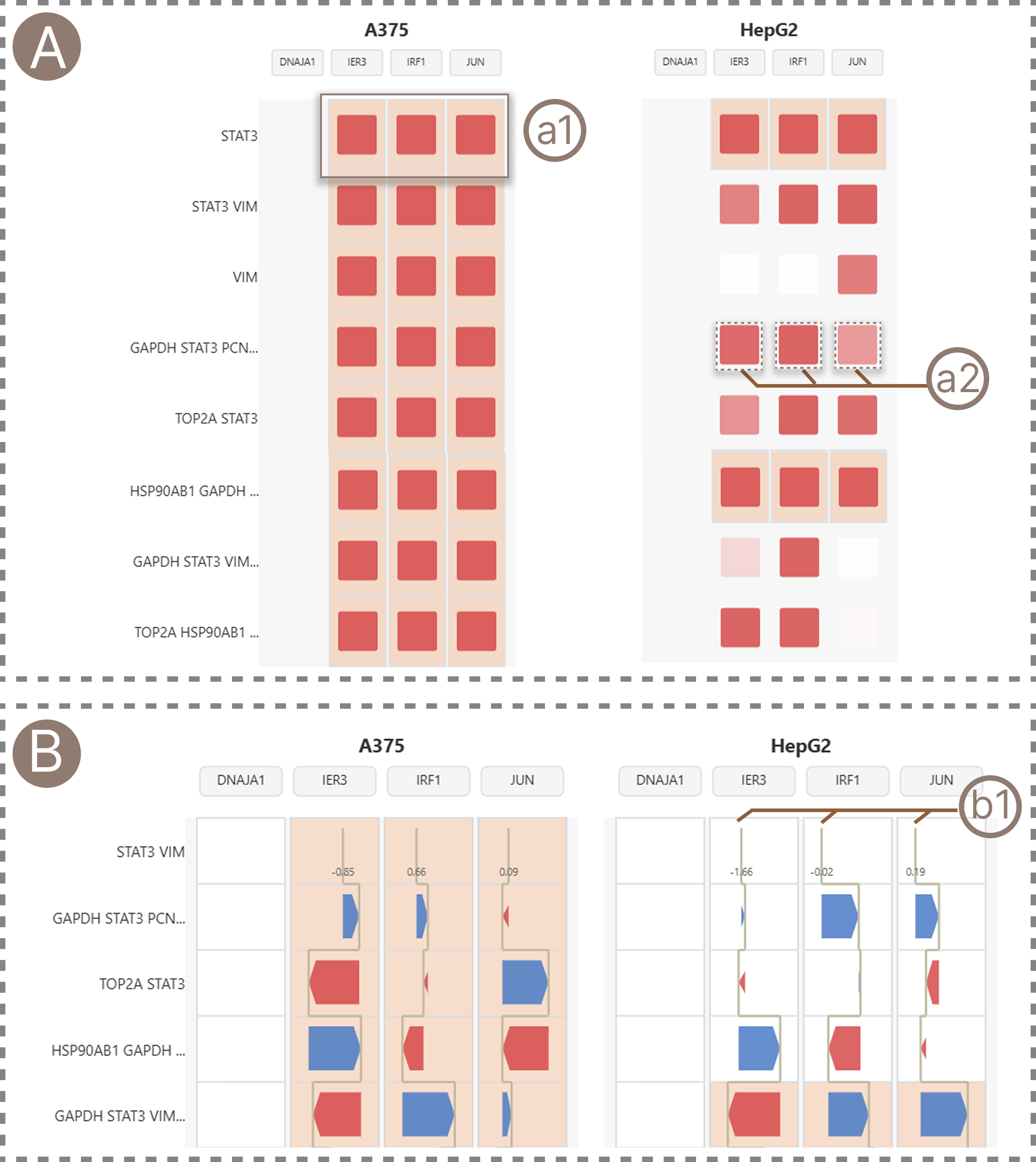}
    \caption{Two complementary modes of the Expression Comparison View. (A) Screening Mode provides a compact overview for rapidly filtering perturbations based on their overall expression patterns relative to filtering criteria predefined by domain experts. (B) Comparison Mode supports fine-grained, gene-level comparison of expression differences across perturbation strategies using directional arrows.}
    \label{fig:perturbation_cell_gene relationship}
\end{figure}

\sidecomment{R2C7}
\textit{\textbf{Screening Mode.}}
Screening Mode (\cref{fig:perturbation_cell_gene relationship}-A) is designed for the rapid elimination of ineffective perturbations from a set of potential candidates. Its primary goal is to help experts quickly assess whether cells satisfy the preset filter threshold requirements.
We adopt a heatmap-based representation, where color saturation encodes the distance between a gene's expression level and the filtering criteria. Darker colors indicate closer proximity to the critical point (\cref{fig:perturbation_cell_gene relationship}-a2). \revise{When a gene's expression value satisfies the preset threshold, the background of the corresponding cell is highlighted (\cref{fig:perturbation_cell_gene relationship}-a1).} This encoding enables experts to scan across multiple perturbation strategies and immediately discard those that fail to induce sufficient expression shifts.

\textit{\textbf{Comparison Mode.}} 
After narrowing down candidates, experts switch to Comparison Mode (\cref{fig:perturbation_cell_gene relationship}-B) to conduct detailed, gene-level comparisons across perturbation strategies. This mode reveals subtle expression differences of the same gene under different perturbations.
\sidecomment{R2C19, R1C1}
\revise{
For each gene column, the vertical guide indicates the predicted expression value of the current perturbation strategy (\cref{fig:perturbation_cell_gene relationship}-b1). The directional arrow is anchored to this value guide and encodes how the expression changes from the perturbation strategy in the previous row to the current one: the arrow direction indicates whether the expression increases or decreases, and the arrow length represents the magnitude of this change. This design allows experts to directly compare expression trajectories of individual genes across multiple perturbations and identify subtle yet meaningful deviations to be identified at a glance. As in Screening Mode, when a gene's expression value satisfies the preset threshold, the background of the corresponding cell is highlighted.}

\textbf{Interactions.}
Several interaction mechanisms support in-depth comparative analysis. When an expert selects a perturbation in the Cluster Navigation View, the system automatically retrieves its five nearest neighbors based on spatial similarity, along with all functional subsets of that perturbation. This allows experts to explore similar strategies and assess their differences.
\sidecomment{R1C1}
To further refine comparisons, experts can manually reorder matrix rows to align perturbations of interest. 
\revise{A filter bar enables dynamic adjustment of the filtering criteria, automatically highlighting perturbations that meet the user-defined criteria.} 
Finally, the view supports manual injection of perturbations, allowing experts to test hypotheses and conduct confirmatory analyses.

\subsection{Collection View}
\sidecomment{R1C2,R2C21,R3C1}
\revise{The Collection View (\cref{fig:teaser}-D) maintains a curated repository of perturbation strategies the user selected during exploration to support final decision-making. It enables users to perform a side-by-side comparison of all discovered and selected promising candidates against each other, evaluating their multi-objective trade-offs to make the final decision on which lead candidate to advance (\cref{fig:main}-B4). For each cell type, we use area charts to present the outcomes of these collected perturbations
\sidecomment{R2C7} \revise{(DR5)}. The horizontal axis encodes the sequence of collected perturbations, while the vertical axis indicates differential gene expression.} Since experts primarily reason about whether genes promote or inhibit cell state, we adopt a unified color scheme—blue for positive regulators and red for negative regulators. Compared to dense multi-line plots, this area-based encoding reduces visual clutter and cognitive load, allowing experts to quickly assess overall trends and trade-offs across the selected perturbation strategies.
\sidecomment{R2C22}
\revise{To ensure scalability during long analysis sessions with a large number of candidates, the Collection View uses a compact area-based encoding, so that multiple collected perturbation strategies can be compressed along the horizontal axis without requiring a large space. For longer sessions, the view also supports horizontal scrolling to accommodate more collected strategies. In addition, experts can actively prune the repository by removing less promising strategies via their axis labels to keep the view focused. 
}

\section{Case Study}

In tumor precision therapy, a central challenge in drug development is identifying treatment regimens that can effectively target multiple cancer cell lines while minimizing toxicity to normal tissues. We present the exploratory process of E1, a domain expert who participated in the iterative design of our system, as a case study to illustrate how \tool supports the discovery of a potential universal anti-tumor intervention strategy across different tissue types. E1's primary objective is to identify precise perturbations that induce desired responses in \texttt{Cell-A375}. In parallel, he evaluates whether the same regimen exhibits consistent efficacy against liver cancer (\texttt{Cell-HepG2}) and leukemia (\texttt{Cell-K562}). Critically, the candidate regimen must maintain the baseline functional state of human induced pluripotent stem cells (\texttt{Cell-iPSC}), thereby ensuring the selectivity of the perturbation strategy.
We integrated several cell datasets from two public repositories, scPerturb \footnote{\url{https://projects.sanderlab.org/scperturb/}} and Zenodo \footnote{\url{https://zenodo.org/records/13350497}}, to support the expert exploration requirements.
\sidecomment{R2C23} 
\revise{These datasets provide gene expression profiles for the target cells and protected cells considered in the case study. We used PerturbNet \cite{yu2025perturbnet} – one of the most widely used virtual cell models – to predict differential gene expression under candidate gene perturbation strategies.}
 This high-precision virtual cell simulation provides a reliable data foundation for expert exploration, supporting interactive screening and analysis within complex perturbation spaces.

\begin{figure}[h]
    \centering
    \includegraphics[width=\linewidth]{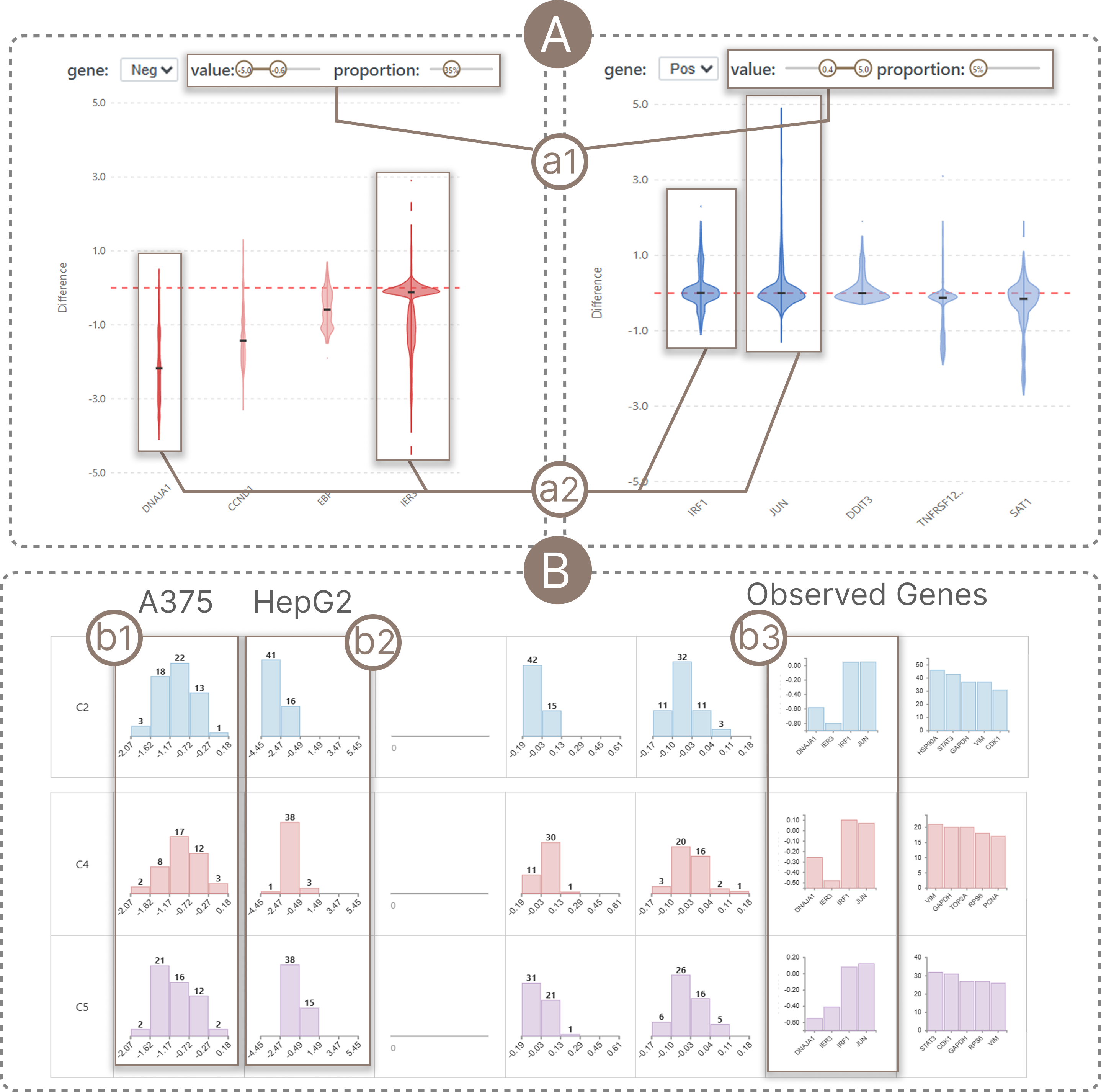}
    \caption{(A) Experts prioritize genes by adjusting filtering parameters (a1) to focus on a subset of four candidate genes (a2). (B) Further screen the clusters. The gene expression contribution of IER3 in Cell-A375 and Cell-HepG2 of  cluster\_4 was generally smaller (b1-b2).}
    \label{fig:case1}
\end{figure}

\textbf{Identify potential observed genes (DR1).} 
E1 began by narrowing down a large pool of function-relevant genes to a manageable set of candidates. Drawing on his domain experience, E1 adopted a pragmatic heuristic often used in exploratory screening: genes with more extreme expression differences are prioritized, as they tend to provide clearer signals and are easier to validate in early-stage analysis.
Using the violin-based filtering component (\cref{fig:case1}-a1), E1 interactively removed genes with weak or ambiguous expression difference patterns. For inhibitory genes, he screened out four negative regulatory genes (\cref{fig:case1}-A, left). Applying a similar logic to positive regulatory genes, he filtered five positive regulatory genes (\cref{fig:case1}-A, right). 
After a comparative review, E1 further refined the list by prioritizing genes with the most extreme values (\cref{fig:case1}-a2). 
\textit{``This may indicate that these genes undergo larger perturbation-induced expression shifts and therefore warrant closer examination.''}
Consequently, he selected two negative regulatory genes (\texttt{Gene-DNAJA1} and \texttt{Gene-IER3}) with the lowest minimum expression, reasoning that stronger suppression effects are more likely to meet his requirements. In parallel, he chose the two highest-ranked positive regulatory genes (\texttt{Gene-IRF1} and \texttt{Gene-JUN}) as representative drivers of the desired cellular state.

\textbf{Locating target clusters within the perturbation space (DR2).} 
With candidate genes identified, E1 shifted focus to exploring the perturbation space in the Cluster Navigation View. Initially, the system presented a large number of fine-grained micro-clusters, making it difficult to reason about global patterns. To address this, E1 adjusted the clustering distance, allowing the system to merge perturbations into eight higher-level clusters.
To quickly assess which clusters were most promising, E1 examined the Observed Genes column in the Cluster Overview (\cref{fig:case1}-b3). He identified cluster\_2, cluster\_4, and cluster\_5 as particularly well aligned with his analytical objective, as they shared a consistent expression pattern: low levels of negative regulatory genes combined with high levels of positive regulatory genes.
This configuration aligned with his expectation of a cell state for cancer cells.
Despite this narrowing, each cluster still contained a large number of perturbations. To further refine the search, E1 focused specifically on the target \texttt{Cell-A375} cells and compared gene expression distributions across the three candidate clusters (\cref{fig:case1}-b1). He observed that cluster\_2 exhibited relatively high \texttt{Gene-IER3} expression, which conflicted with the desired suppression pattern, as \texttt{Gene-IER3} is a negative regulatory gene whose elevated expression may counteract the intended cellular response. In contrast, cluster\_4 and cluster\_5 more closely matched the expected cellular response.
Between these two, cluster\_4 showed a broader distribution of lower gene expression in \texttt{Cell-HepG2}, indicating a more consistent response across different target cell lines. Based on this observation, E1 prioritized cluster\_4 for further analysis and temporarily hid the remaining clusters to reduce visual clutter.

\begin{figure}[h]
    \centering
    \includegraphics[width=0.8\linewidth]{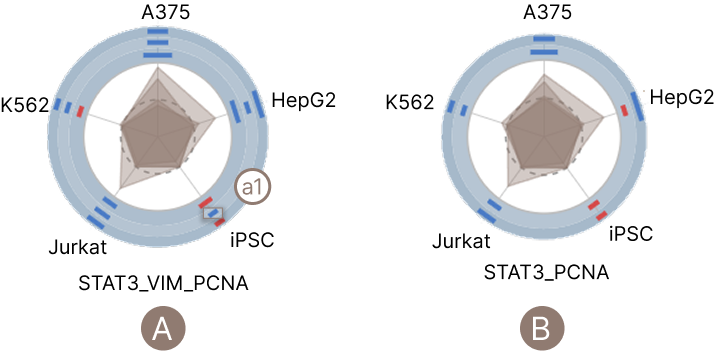}
    \caption{(A) Candidate perturbation from initial exploration. (B) Target perturbation precisely localized through SHAP contribution analysis.}
    \label{fig:case2}
\end{figure}

\textbf{Understand the contribution of perturbed genes to cell profiles (DR3).} 
After zooming into cluster\_4, the system transitioned to a radar-based glyph representation, enabling direct comparison of individual perturbations across different cell lines. During this exploration, E1 noticed that the perturbation \texttt{STAT3\_VIM\_PCNA} (\cref{fig:case2}-A), which represents the simultaneous perturbation of three genes (\texttt{Gene-STAT3}, \texttt{Gene-VIM}, and \texttt{Gene-PCNA}), produced strong gene expression shifts in both \texttt{Cell-A375} and \texttt{Cell-HepG2}.
However, the same perturbation also caused noticeable deviations in protected \texttt{Cell-iPSC}, raising safety concerns.
To understand the source of this undesired effect, E1 examined the SHAP-based contribution blocks surrounding the glyph (\cref{fig:case2}-A). These revealed that \texttt{Gene-VIM} contributed disproportionately to gene expression differences in \texttt{Cell-iPSC}, as indicated by the middle ring of the outer contribution encoding in \cref{fig:case2}-a1. Based on this insight, E1 hypothesized that removing \texttt{Gene-VIM} from the perturbation might preserve anti-tumor efficacy while reducing harm to normal cells.
Therefore, he used the search function to locate perturbations excluding \texttt{Gene-VIM}, focusing on \texttt{STAT3\_PCNA}. After inspection, this perturbation appeared promising (\cref{fig:case2}-B), prompting E1 to initiate a more detailed analysis by selecting it for in-depth comparison.

\begin{figure}[h]
    \centering
    \includegraphics[width=\linewidth]{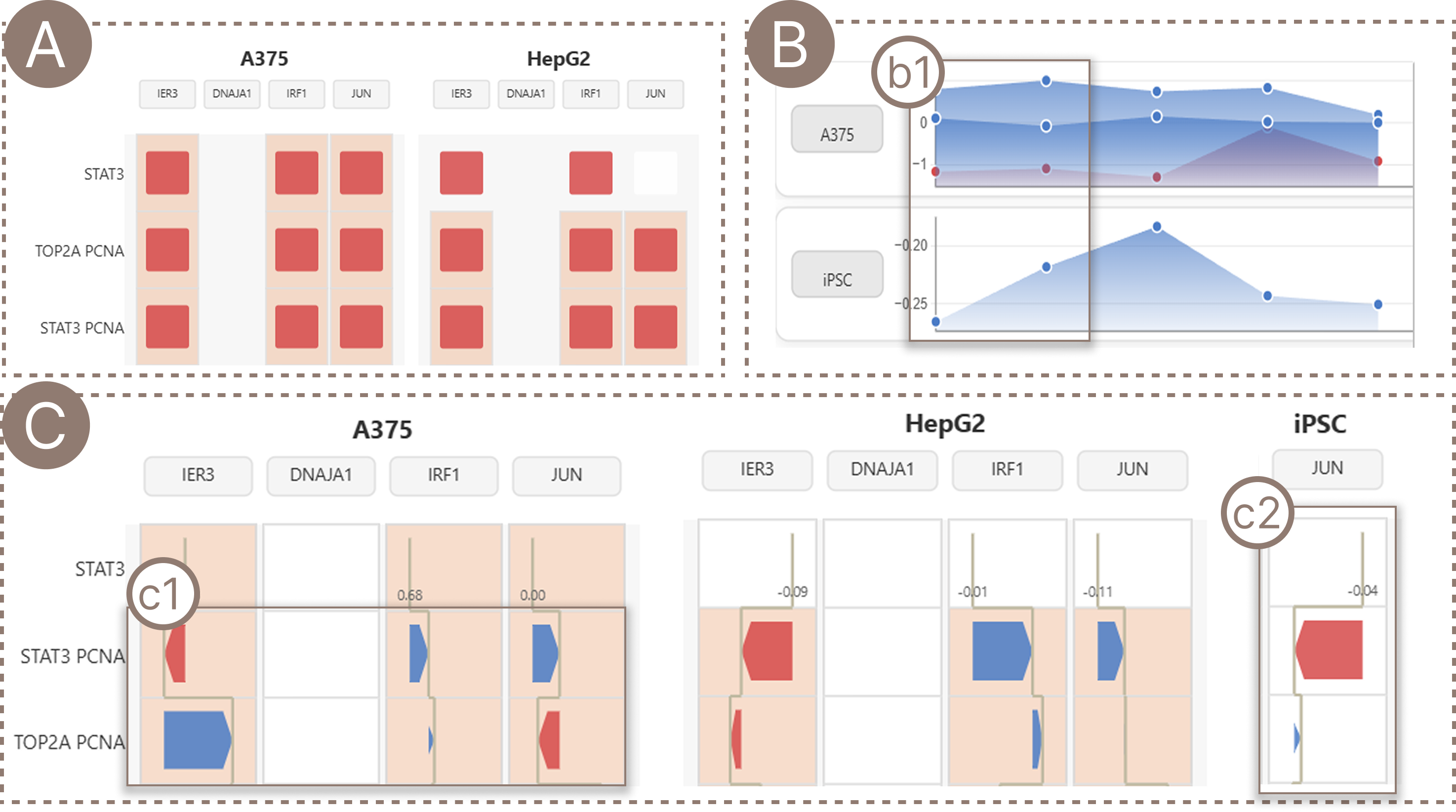}
    \caption{(A) Experts filtered perturbations based on predefined criteria. (B) The Collection View facilitates trade-off analysis between efficacy and side effects. (C) Detailed analysis of the direction of gene expression changes between perturbations assists in identifying perturbations that better align with the desired regulatory objectives.}
    \label{fig:case3}
\end{figure}

\textbf{In-depth comparison of potential perturbation candidates (DR4).}
\sidecomment{R2C24}
\revise{In the Expression Comparison View, the system automatically constructed a comparison set of eight candidate perturbation strategies for initial screening, including the five nearest neighbors of \texttt{STAT3\_PCNA} based on the spatial distance in the cluster projection, along with subsets derived from the selected perturbation combination (e.g., \texttt{STAT3}, \texttt{PCNA}). He then explored various
perturbation strategies that exhibited gene expression patterns similar
to those of \texttt{STAT3\_PCNA}. }
Based on the similarity comparison, he successfully identified several perturbation strategies with comparable expression patterns, such as \texttt{STAT3} and \texttt{TOP2A\_PCNA}.
Leveraging his prior knowledge, he adjusted the filtering criteria to define a more stringent standard for achieving the desired cellular response.
Based on the color saturation in the heatmap (\cref{fig:case3}-A), E1 rapidly identified perturbations that satisfied these criteria. Both \texttt{STAT3\_PCNA} and \texttt{TOP2A\_PCNA} successfully triggered the desired expression differences in \texttt{Cell-A375} and \texttt{Cell-HepG2}. To compare them more precisely, he reordered the perturbations and entered the detailed comparison mode (\cref{fig:case3}-C).
This view revealed that \texttt{STAT3\_PCNA} exhibited a more favorable regulatory profile in \texttt{Cell-A375} (\cref{fig:case3}-c1), characterized by stronger suppression of the negative regulatory gene (\texttt{IER3}) and higher activation of the positive regulatory gene (\texttt{JUN}). More importantly, the gene expression difference in \texttt{Cell-iPSC} was further from the filtering criteria with no visual background highlighting, indicating improved safety (\cref{fig:case3}-c2). Therefore, E1 marked \texttt{STAT3\_PCNA} as a high-potential candidate and added it to the Collection View for later comparison.

\textbf{Trade-off analysis and final candidate selection (DR5).} Over multiple iterations, the Collection View accumulated five representative perturbation candidates:
\texttt{STAT3\_PCNA}, \texttt{TOP2A\_VIM\_PCNA}, \texttt{TOP2A\_STAT3}, \texttt{TOP2A\_PCNA}, and \texttt{VIM\_RPS6} (\cref{fig:case3}-B).
Given the redundancy observed in earlier comparisons, E1 removed \texttt{TOP2A\_PCNA} and conducted a final side-by-side evaluation of the remaining candidates. He observed that although \texttt{TOP2A\_VIM\_PCNA} achieved the most significant expression shifts on \texttt{Cell-A375}, it also posed higher risks to \texttt{Cell-iPSC} (\cref{fig:case3}-b1). In contrast, \texttt{STAT3\_PCNA} offered the best balance between target efficacy and safety, with only a marginal reduction in tumor-killing power.
Based on this comprehensive trade-off analysis, E1 selected \texttt{Gene-STAT3} and \texttt{Gene-PCNA} as the outcome of this visual analytics session and prepared them for subsequent wet-lab validation.
\section{Expert Interview}
We conducted semi-structured expert interviews to evaluate the effectiveness, usability, and practical value of \tool in realistic analysis scenarios.
\revise{After obtaining approval from the School of Computer Science and Engineering, Southeast University,} we recruited three domain experts (P1–P3) working in virtual cell modeling and drug discovery.
None of the experts had been involved in the design or development of our system. P1 is a research assistant with three years of research experience, while P2 and P3 are Ph.D. candidates with more than four years of relevant research experience in computational biology and related areas. All participants were familiar with gene perturbation analysis workflows and routinely conducted exploratory biological analyses.

Each interview session lasted approximately 85 minutes and followed a four-stage protocol. \revise{After obtaining the consent of participants, we first introduced the research background, core analytical goals, and the design rationale of each view in \tool (15 minutes) to familiarize
them with the task}. We then demonstrated the system workflow through a concrete perturbation analysis example (15 minutes). Next, each expert was given 25 minutes of hands-on exploration time.
Experts were encouraged to think aloud during the exploration.
This enabled us to capture their in-situ observations, reasoning processes, and confusions as they interacted with the system. Finally, we conducted a 30-minute semi-structured interview focusing on the overall workflow, visual encodings, interaction mechanisms, perceived strengths and limitations, and potential future extensions.
The interview data were analyzed through thematic analysis~\cite{clarke2017thematic}.

\textbf{System Workflow.} 
All experts appreciated \tool's overall workflow. 
They noted that, in the traditional perturbation discovery process, researchers often need to manually enumerate candidate perturbation combinations and sequentially input them into predictive models to observe the outcomes. 
This process is not only time-consuming but also makes it difficult to form a comprehensive understanding of the global perturbation space or to reason about relationships between different perturbation strategies.
In contrast, the experts highlighted that \tool bridges the gap between exploring global perturbation patterns and local mechanisms through structured, interactive visual exploration that systematically orchestrates perturbation discovery.
The design choices of visually summarizing perturbation strategies via the cluster projection plot and providing detailed SHAP-based contribution explanations were regarded by experts as ``efficient'', ``intuitive'', and ``insightful''. 
As P1 noted, \textit{``This way of exploring perturbations as a space rather than isolated inputs is very powerful. It avoids the inefficiency of testing strategies one by one and allows us to compare different perturbation strategies systematically, which is something we usually struggle with.''}
The experts also appreciated the system's redirection and navigation mechanisms, which fit into their highly iterative hypothesis-validation workflows.
In particular, the ability to directly search for and locate specific perturbations based on prior hypotheses or newly discovered clues was considered highly valuable for hypothesis-driven validation. 
Importantly, P1 noted that the workflow is not limited to the task we focused on in this paper: by replacing the observed gene set with genes associated with different biological functions, the same workflow could be readily generalized to other biological tasks, such as anti-proliferative drug discovery or cancer treatment design.

\textbf{Visual Design and Interaction.} 
The experts unanimously agreed that \tool presents a well-structured interface and effectively prioritizes critical information needed for perturbation analysis. They emphasized that the system balances overview and detail, allowing users to transition smoothly between coarse-grained screening and fine-grained inspection.
Furthermore, experts are highly commended the radar-based glyph design.
They highlighted its effectiveness in aggregating diverse observed gene values across multiple cell lines into a unified geometric representation.
In particular, P3 emphasized that mapping independent gene axes into a single polygonal outline allows them to quickly perceive holistic expression patterns.
P1 also stated, \textit{``This design enables intuitive understanding of the relationships among multiple perturbed genes, observed genes, and cell populations, allowing us to focus on biologically meaningful patterns rather than low-level data parsing.''}
Regarding interaction, P2 appreciated the semantic zoom mechanism, which enables a fluid transition from global cluster distributions to localized functional details by progressively manifesting data points as detailed glyphs. Furthermore, P2 also praised the integration of SHAP values within the glyph’s outer arcs, which explicitly encodes the magnitude and direction of gene contributions to render the underlying biological mechanisms transparent. He remarked that this interpretability provides a heuristic guide that empowers them to purposefully navigate the perturbation space and optimize strategies for subsequent rounds of combinatorial perturbations.

\textbf{Suggestions.} 
Experts also offered valuable suggestions for improving \tool.
First, P3 pointed out that our system could integrate LLMs to recommend perturbations tailored to users' specific experimental needs. 
However, he noted that LLM recommendations alone are insufficient to navigate the complexity of early-stage exploration.
Rather than undermining the utility of the Expression Comparison View, such integration would further underscore its importance:  \textit{``Even with LLM recommendations, since analysis goals in early exploration are often vague and potential side effects remain unclear, we still need visual comparison and exploration to finalize the most suitable perturbation strategy.''} 
Second, expert P1 mentioned that beyond gene expression levels, the positional attributes of genes within cells also carry significant biological meaning. 
Future iterations of the system could consider integrating more metrics to provide richer visual analysis capabilities.
\section{Discussion}
In this section, we reflect on broader insights arising from the design and evaluation of \tool, discuss the generalizability and boundaries of our approach, and outline limitations and directions for future work.

\textbf{Design implications.}  
Based on our iterative design process and expert feedback, we distill two design implications from this study that may inform future visual analytics systems for biological data and, more broadly, for AI-assisted scientific exploration.

\begin{itemize}[leftmargin=*, topsep=4pt, itemsep=2pt]

\item \textit{Use hierarchical exploration to match expert reasoning granularity.} Our two-level exploration strategy—first clustering perturbations to identify promising regions, then drilling down into individual perturbations—was not driven solely by data size, but by the way experts naturally reason under uncertainty. Experts initially reason at a coarse level, asking questions such as ``which types of perturbations tend to work?'', before committing attention to detailed comparisons among specific candidates.
Directly exposing fine-grained perturbations from the outset would overload users and obscure higher-level regularities. The hierarchical design allows experts to progressively align analytical granularity with decision confidence, narrowing the search space only when sufficient evidence accumulates.

\item \textit{Use feature contribution as a retrieval and pruning mechanism.} Feature contribution techniques (e.g., SHAP) are used to explain model predictions by attributing outcomes to individual features. Our study suggests an alternative and complementary design usage: treating feature contributions as interactive retrieval cues for exploration and pruning, rather than as definitive explanations. In our system, contribution patterns are used to identify which components within a multi-gene perturbation disproportionately influence specific cell types, allowing experts to search for and compare related perturbations that exclude or modify those components.
This design supports expert reasoning through elimination and refinement, enabling users to narrow the perturbation space without requiring them to fully trust or interpret the underlying model explanations. Designers of visual analytics systems may consider framing feature contribution not only as an explanatory output, but also as an operational mechanism for search, comparison, and hypothesis pruning in exploratory analysis.
\end{itemize}

\textbf{Generalizability.} 
Although our case study is grounded in perturbation-based cell analysis, the generalizability of the system lies in several concrete design mechanisms rather than the specific dataset or biological task. First, the system treats combinatorial actions (e.g., multi-gene perturbations) as primary analytical units. The glyph-based representation explicitly supports many-to-many relationships by simultaneously encoding multiple perturbed genes, multiple observed response features, and multiple target entities (e.g., different cell types) within a single visual object. This design is applicable to other tasks where experts need to reason about compound interventions and their heterogeneous effects across multiple evaluation contexts, without requiring one-to-one alignment between actions and outcomes.
Furthermore, feature contribution is used not as a standalone explanation output, but as a retrieval and pruning cue to locate related variants and eliminate undesirable components. This usage does not depend on specific biological assumptions and can be reused in other model-driven exploration tasks where understanding which part of an action to modify or remove is more important than fully interpreting the model.

\sidecomment{Second Round revised}
\textbf{Limitation and future work.}
\revise{We identify two limitations of the current system and discuss future directions.} First, our system assumes that the candidate perturbation space is pre-generated by upstream computational or experimental methods. Users are required to provide the perturbed genes they want to test, and the system automatically generates all combinatorial actions based on this input. Consequently, the system does not explore novel perturbations outside the user-specified space, which may limit the discovery of unexpected or rare perturbation effects. In the future, this limitation could be addressed by integrating methods for suggesting promising perturbations beyond the initial user-specified set, such as predictive models or active learning strategies that prioritize unexplored combinations.
Second, the system currently does not incorporate a mechanism for signaling prediction uncertainty. Although our models achieve high accuracy in the current study, in scenarios where model performance may be lower or the biological effects are more complex, additional mechanisms would be needed to communicate uncertainty to the user. This could include confidence intervals, probabilistic scores, or interactive visual cues that help users interpret and weigh the reliability of the system’s predictions.

\section{Conclusion}
In this work, we present \tool to support the exploration of gene perturbation strategies for cell–type–specific drug discovery and side-effect assessment. \tool organizes high-dimensional perturbation outcomes through clustering-based overviews, enabling researchers to identify groups of perturbations with similar gene expression patterns. To facilitate comparison and interpretation of perturbation strategies at the expression level, the system further employs a glyph-based representation that summarizes gene expression responses across cell types for each perturbation. These representations are integrated within coordinated views to support interactive exploration from a global overview to detailed analysis. A case study and expert interviews demonstrate the system's usability and its effectiveness.
Finally, we envision future work enhancing \tool by suggesting novel perturbations, conveying prediction uncertainty, and further supporting exploratory decision-making in complex biological contexts.

\acknowledgments{%
This work is supported in part by the Natural Science Foundation
of Jiangsu Province, China (Grant No.BK20241300) and in part by the Young Scientists Fund of the National Natural Science Foundation of China (Grant No.62502087).
We also thank the anonymous reviewers for their valuable comments and constructive suggestions.
}

\bibliographystyle{abbrv-doi-hyperref}
\bibliography{main}

@String{Computing = "Computing" }

@String{Computer = "{IEEE} Computer" }

@String{Springer = "Springer-Verlag" }

@article{clarke2017thematic,
  title={{Thematic Analysis}},
  author={Clarke, Victoria and Braun, Virginia},
  journal={The Journal of Positive Psychology},
  volume={12},
  number={3},
  pages={297--298},
  year={2017},
  publisher={Taylor \& Francis},
  doi = {10.1080/17439760.2016.1262613}
}

@article{adduri2025predicting,
  title={{Predicting Cellular Responses to Perturbation Across Diverse Contexts with State}},
  author={Adduri, Abhinav K and Gautam, Dhruv and Bevilacqua, Beatrice and Imran, Alishba and Shah, Rohan and Naghipourfar, Mohsen and Teyssier, Noam and Ilango, Rajesh and Nagaraj, Sanjay and Dong, Mingze and others},
  journal={bioRxiv},
  year={2025},
  publisher={Cold Spring Harbor Laboratory},
  doi={10.1101/2025.06.26.661135}
}

@article{kalter2025off,
  address = {United States},
	title = {{Off-target Effects in {CRISPR}-{Cas} Genome Editing for Human Therapeutics: {Progress} and Challenges}},
	volume = {36},
	issn = {2162-2531},
	doi = {10.1016/j.omtn.2025.102636},
	language = {eng},
	number = {3},
	journal = {Molecular Therapy Nucleic Acids},
	author = {Kalter, Nechama and Fuster-García, Carla and Silva, Alfredo and Ronco-Díaz, Víctor and Roncelli, Stefano and Turchiano, Giandomenico and Gorodkin, Jan and Cathomen, Toni and Benabdellah, Karim and Lee, Ciaran and Hendel, Ayal},
	year = {2025},
	pages = {102636},
}

@article{khikhmetova2025advances,
  title={{Advances in Drug Discovery: Navigating Challenges and Embracing Innovation}},
  author={Khikhmetova, Kamila},
  journal={Australian Journal of Biomedical Research},
  volume={1},
  number={1},
  pages={aubm005},
  year={2025},
  publisher={Australasia Publishing Group},
  doi={10.63946/aubiomed/16813}
}

@article{yu2025perturbnet,
  title={{PerturbNet Predicts Single-cell Responses to Unseen Chemical and Genetic Perturbations}},
  author={Yu, Hengshi and Qian, Weizhou and Song, Yuxuan and Welch, Joshua D},
  journal={Molecular Systems Biology},
  volume={21},
  number={8},
  pages={960--982},
  year={2025},
  doi={10.1038/s44320-025-00131-3}
}

@article{tang2025cellforge,
  title={{CellForge: Agentic Design of Virtual Cell Models}},
  author={Tang, Xiangru and Yu, Zhuoyun and Chen, Jiapeng and Cui, Yan and Shao, Daniel and Wang, Weixu and Wu, Fang and Zhuang, Yuchen and Shi, Wenqi and Huang, Zhi and others},
  journal={arXiv preprint arXiv:2508.02276},
  year={2025},
  doi={10.48550/arXiv.2508.02276}
}

@article{klein2025cellflow,
  title={{CellFlow Enables Generative Single-Cell Phenotype Modeling with Flow Matching}},
  author={Klein, Dominik and Fleck, Jonas Simon and Bobrovskiy, Daniil and Zimmermann, Lea and Becker, S{\"o}ren and Palma, Alessandro and Dony, Leander and Tejada-Lapuerta, Alejandro and Huguet, Guillaume and Lin, Hsiu-Chuan and others},
  journal={bioRxiv},
  year={2025},
  publisher={Cold Spring Harbor Laboratory},
  doi={10.1101/2025.04.11.648220}
}

@article{lotfollahi2019scgen,
  title={{scGen Predicts Single-Cell Perturbation Responses}},
  author={Lotfollahi, Mohammad and Wolf, F Alexander and Theis, Fabian J},
  journal={Nature Methods},
  volume={16},
  number={8},
  pages={715--721},
  year={2019},
  publisher={Nature Publishing Group US New York},
  doi={10.1038/s41592-019-0494-8}
}

@article{simon2017visexpress,
  title={{VisExpress: Visual Exploration of Differential Gene Expression Data}},
  author={Simon, Svenja and Mittelst{\"a}dt, Sebastian and Kwon, Bum Chul and Stoffel, Andreas and Landstorfer, Richard and Neuhaus, Klaus and M{\"u}hlig, Anna and Scherer, Siegfried and Keim, Daniel A},
  journal={Information Visualization},
  volume={16},
  number={1},
  pages={48--73},
  year={2017},
  publisher={SAGE Publications Sage UK: London, England},
  doi = {10.1177/1473871615612883}
}

@article{patil2023scviewer,
  title={{scViewer: An Interactive Single-Cell Gene Expression Visualization Tool}},
  author={Patil, Abhijeet R and Kumar, Gaurav and Zhou, Huanyu and Warren, Liling},
  journal={Cells},
  volume={12},
  number={11},
  pages={1489},
  year={2023},
  publisher={MDPI},
  doi = {10.3390/cells12111489}
}

@ARTICLE{Sheng2026CellScout,
  author={Sheng, Rui and Zang, Zelin and Wang, Jiachen and Luo, Yan and Chen, Zixin and Zhou, Yan and Ruan, Shaolun and Qu, Huamin},
  journal={IEEE Transactions on Visualization and Computer Graphics}, 
  title={{CellScout: Visual Analytics for Mining Biomarkers in Cell State Discovery}}, 
  year={2026},
  volume={32},
  number={2},
  pages={1497-1512},
  doi={10.1109/TVCG.2025.3636102}
}

@article{wei2025perturbase,
  title={{PerturBase: A Comprehensive Database for Single-Cell Perturbation Data Analysis and Visualization}},
  author={Wei, Zhiting and Si, Duanmiao and Duan, Bin and Gao, Yicheng and Yu, Qian and Zhang, Zhenbo and Guo, Ling and Liu, Qi},
  journal={Nucleic Acids Research},
  volume={53},
  number={D1},
  pages={D1099--D1111},
  year={2025},
  publisher={Oxford University Press},
  doi={10.1093/nar/gkae858}
}

@article{zhou2019networkanalyst,
  title={{NetworkAnalyst 3.0: A Visual Analytics Platform for Comprehensive Gene Expression Profiling and Meta-Analysis}},
  author={Zhou, Guangyan and Soufan, Othman and Ewald, Jessica and Hancock, Robert EW and Basu, Niladri and Xia, Jianguo},
  journal={Nucleic Acids Research},
  volume={47},
  number={W1},
  pages={W234--W241},
  year={2019},
  publisher={Oxford University Press},
  doi={https://doi.org/10.1093/nar/gkz240}
}

@article{emmerich2021improving,
  title={{Improving Target Assessment in Biomedical Research: the GOT-IT Recommendations}},
  author={Emmerich, Christoph H and Gamboa, Lorena Martinez and Hofmann, Martine CJ and Bonin-Andresen, Marc and Arbach, Olga and Schendel, Pascal and Gerlach, Bj{\"o}rn and Hempel, Katja and Bespalov, Anton and Dirnagl, Ulrich and others},
  journal={Nature Reviews Drug Discovery},
  volume={20},
  number={1},
  pages={64--81},
  year={2021},
  publisher={Nature Publishing Group UK London},
  doi={10.1038/s41573-020-0087-3}
}

@article{wong2021search,
  title={{Search and Visualization of Gene-Drug-Disease Interactions for Pharmacogenomics and Precision Medicine Research Using GeneDive}},
  author={Wong, Mike and Previde, Paul and Cole, Jack and Thomas, Brook and Laxmeshwar, Nayana and Mallory, Emily and Lever, Jake and Petkovic, Dragutin and Altman, Russ B and Kulkarni, Anagha},
  journal={Journal of Biomedical Informatics},
  volume={117},
  pages={103732},
  year={2021},
  publisher={Elsevier},
  doi={https://doi.org/10.1016/j.jbi.2021.103732}
}

@article{li2022deep,
  title={{Deep Learning Methods for Molecular Representation and Property Prediction}},
  author={Li, Zhen and Jiang, Mingjian and Wang, Shuang and Zhang, Shugang},
  journal={Drug Discovery Today},
  volume={27},
  number={12},
  pages={103373},
  year={2022},
  publisher={Elsevier},
  doi={https://doi.org/10.1016/j.drudis.2022.103373}
}

@article{liu2024vibrant,
  title={{VIBRANT: Spectral Profiling for Single-Cell Drug Responses}},
  author={Liu, Xinwen and Shi, Lixue and Zhao, Zhilun and Shu, Jian and Min, Wei},
  journal={Nature Methods},
  volume={21},
  number={3},
  pages={501--511},
  year={2024},
  publisher={Nature Publishing Group US New York},
  doi={10.1038/s41592-024-02185-x}
}

@article{bunne2024build,
  title={{How to Build the Virtual Cell with Artificial Intelligence: Priorities and Opportunities}},
  author={Bunne, Charlotte and Roohani, Yusuf and Rosen, Yanay and Gupta, Ankit and Zhang, Xikun and Roed, Marcel and Alexandrov, Theo and AlQuraishi, Mohammed and Brennan, Patricia and Burkhardt, Daniel B and others},
  journal={Cell},
  volume={187},
  number={25},
  pages={7045--7063},
  year={2024},
  publisher={Elsevier},
  doi={10.1016/j.cell.2024.11.015}
}

@article{fang2025cell,
  title={{Cell-GraphCompass: Modeling Single Cells with Graph Structure Foundation Model}},
  author={Fang, Chen and Cui, Wentao and Hu, Zhilong and Liu, Wenhao and Chen, Shubai and Chang, Shaole and Long, Qingqing and Li, Cong and Liu, Yana and Jiang, Haiping and others},
  journal={National Science Review},
  volume={12},
  number={10},
  pages={nwaf255},
  year={2025},
  publisher={Oxford University Press},
  doi={10.1093/nsr/nwaf255}
}

@article{roohani2024predicting,
  title={{Predicting Transcriptional Outcomes of Novel Multigene Perturbations with GEARS}},
  author={Roohani, Yusuf and Huang, Kexin and Leskovec, Jure},
  journal={Nature Biotechnology},
  volume={42},
  number={6},
  pages={927--935},
  year={2024},
  publisher={Nature Publishing Group US New York},
  doi={10.1038/s41587-023-01905-6}
}

@article{ma2025ai,
  title={{AI-Driven Virtual Cell Models in Preclinical Research: Technical Pathways, Validation Mechanisms, and Clinical Translation Potential}},
  author={Ma, Chunyu and Zhang, Han and Rao, Yiwei and Jiang, Xinyu and Liu, Boheng and Sun, Zhikang and Song, Zhenyu and Gao, Yuan and Cui, Yuhao and Liu, Xinyu and others},
  journal={npj Digital Medicine},
  year={2026},
  volume={9},
  pages={25},
  publisher={Nature Publishing Group UK London},
  doi={10.1038/s41746-025-02198-6}
}

@article{kleverov2024phantasus,
  title={{Phantasus, A Web Application for Visual and Interactive Gene Expression Analysis}},
  author={Kleverov, Maksim and Zenkova, Daria and Kamenev, Vladislav and Sablina, Margarita and Artyomov, Maxim N and Sergushichev, Alexey A},
  journal={eLife},
  volume={13},
  pages={e85722},
  year={2024},
  publisher={eLife Sciences Publications, Ltd},
  doi={10.7554/eLife.85722}
}

@article{megill2021cellxgene,
  title={{Cellxgene: A Performant, Scalable Exploration Platform for High Dimensional Sparse Matrices}},
  author={Megill, Colin and Martin, Bruce and Weaver, Charlotte and Bell, Sidney and Prins, Lia and Badajoz, Seve and McCandless, Brian and Pisco, Angela Oliveira and Kinsella, Marcus and Griffin, Fiona and others},
  journal={bioRxiv},
  year={2021},
  publisher={Cold Spring Harbor Laboratory},
  doi={10.1101/2021.04.05.438318}
}

@article{fernandez2017clustergrammer,
  title={{Clustergrammer, A Web-Based Heatmap Visualization and Analysis Tool for High-dimensional Biological Data}},
  author={Fernandez, Nicolas F and Gundersen, Gregory W and Rahman, Adeeb and Grimes, Mark L and Rikova, Klarisa and Hornbeck, Peter and Ma’ayan, Avi},
  journal={Scientific Data},
  volume={4},
  number={1},
  pages={170151},
  year={2017},
  publisher={Nature Publishing Group},
  doi={10.1038/sdata.2017.151}
}

@article{chen2019single,
  title={{Single-Cell Trajectories Reconstruction, Exploration and Mapping of Omics Data with STREAM}},
  author={Chen, Huidong and Albergante, Luca and Hsu, Jonathan Y and Lareau, Caleb A and Lo Bosco, Giosue and Guan, Jihong and Zhou, Shuigeng and Gorban, Alexander N and Bauer, Daniel E and Aryee, Martin J and others},
  journal={Nature Communications},
  volume={10},
  number={1},
  pages={1903},
  year={2019},
  publisher={Nature Publishing Group UK London},
  doi={10.1038/s41467-019-09670-4}
}

@article{smolander2023cell,
  title={{Cell-Connectivity-Guided Trajectory Inference from Single-Cell Data}},
  author={Smolander, Johannes and Junttila, Sini and Elo, Laura L},
  journal={Bioinformatics},
  volume={39},
  number={9},
  pages={btad515},
  year={2023},
  publisher={Oxford University Press},
  doi={10.1093/bioinformatics/btad515}
}

@article{wong2024splicewiz,
  title={{SpliceWiz: Interactive Analysis and Visualization of Alternative Splicing in R}},
  author={Wong, Alex CH and Wong, Justin JL and Rasko, John EJ and Schmitz, Ulf},
  journal={Briefings in Bioinformatics},
  volume={25},
  number={1},
  pages={bbad468},
  year={2024},
  publisher={Oxford University Press},
  doi={10.1093/bib/bbad468}
}

@article{lex2012stratomex,
author = {Lex, A. and Streit, M. and Schulz, H.-J. and Partl, C. and Schmalstieg, D. and Park, P.J. and Gehlenborg, N.},
title = {{StratomeX: Visual Analysis of Large-Scale Heterogeneous Genomics Data for Cancer Subtype Characterization}},
journal = {Computer Graphics Forum},
volume = {31},
number = {3pt3},
pages = {1175-1184},
doi = {https://doi.org/10.1111/j.1467-8659.2012.03110.x},
year = {2012}
}

@INPROCEEDINGS{roper2023vis,
  author={Roper, Braden and Mathews, James C. and Nadeem, Saad and Park, Ji Hwan},
  booktitle={2023 IEEE Visualization and Visual Analytics (VIS)}, 
  title={{Vis-SPLIT: Interactive Hierarchical Modeling for mRNA Expression Classification}}, 
  year={2023},
  volume={},
  number={},
  publisher={IEEE},
  pages={106-110},
  doi={10.1109/VIS54172.2023.00030},
  address={Melbourne, Australia}
  }

@article{wilson2018automated,
  title={{Automated Literature Mining and Hypothesis Generation Through a Network of Medical Subject Headings}},
  author={Wilson, Stephen Joseph and Wilkins, Angela Dawn and Holt, Matthew V and Choi, Byung Kwon and Konecki, Daniel and Lin, Chih-Hsu and Koire, Amanda and Chen, Yue and Kim, Seon-Young and Wang, Yi and others},
  journal={bioRxiv},
  year={2018},
  publisher={Cold Spring Harbor Laboratory},
  doi={10.1101/403667}
}

@article{cheerkoot2021literature,
  title={{Literature-based Discovery Approaches for Evidence-Based Healthcare: A Systematic Review}},
  author={Cheerkoot-Jalim, Sudha and Khedo, Kavi Kumar},
  journal={Health and Technology},
  volume={11},
  number={6},
  pages={1205--1217},
  year={2021},
  publisher={Springer},
  doi={10.1007/s12553-021-00605-y}
}

@article{istrate2025rbio1,
  title={{rbio1-Training Scientific Reasoning LLMs with Biological World Models as Soft Verifiers}},
  author={Istrate, Ana-Maria and Milletari, Fausto and Castrotorres, Fabrizio and Tomczak, Jakub M and Torkar, Michaela and Li, Donghui and Karaletsos, Theofanis},
  journal={bioRxiv},
  year={2025},
  publisher={Cold Spring Harbor Laboratory},
  doi={10.1101/2025.08.18.670981}
}

@article{wu2018moleculenet,
  title={{MoleculeNet: A Benchmark for Molecular Machine Learning}},
  author={Wu, Zhenqin and Ramsundar, Bharath and Feinberg, Evan N and Gomes, Joseph and Geniesse, Caleb and Pappu, Aneesh S and Leswing, Karl and Pande, Vijay},
  journal={Chemical Science},
  volume={9},
  number={2},
  pages={513--530},
  year={2018},
  publisher={Royal Society of Chemistry},
  doi={10.1039/C7SC02664A}
}

@article{pushpakom2019drug,
  title={{Drug Repurposing: Progress, Challenges and Recommendations}},
  author={Pushpakom, Sudeep and Iorio, Francesco and Eyers, Patrick A and Escott, K Jane and Hopper, Shirley and Wells, Andrew and Doig, Andrew and Guilliams, Tim and Latimer, Joanna and McNamee, Christine and others},
  journal={Nature Reviews Drug Discovery},
  volume={18},
  number={1},
  pages={41--58},
  year={2019},
  publisher={Nature Publishing Group},
  doi={https://doi.org/10.1038/nrd.2018.168}
}

@article{dwivedi2023explainable,
  title={{Explainable AI (XAI): Core Ideas, Techniques, and Solutions}},
  author={Dwivedi, Rudresh and Dave, Devam and Naik, Het and Singhal, Smiti and Omer, Rana and Patel, Pankesh and Qian, Bin and Wen, Zhenyu and Shah, Tejal and Morgan, Graham and others},
  journal={ACM Computing Surveys},
  volume={55},
  number={9},
  year={2023},
  publisher={ACM New York, NY},
  doi = {10.1145/3561048},
  articleno = {194},
  numpages = {33}
}

@article{alicioglu2022survey,
  title={{A Survey of Visual Analytics for Explainable Artificial Intelligence Methods}},
  author={Alicioglu, Gulsum and Sun, Bo},
  journal={Computers \& Graphics},
  volume={102},
  pages={502--520},
  year={2022},
  publisher={Elsevier},
  doi={https://doi.org/10.1016/j.cag.2021.09.002}
}

@article{hughes2011principles,
author = {Hughes, JP and Rees, S and Kalindjian, SB and Philpott, KL},
title = {{Principles of Early Drug Discovery}},
journal = {British Journal of Pharmacology},
volume = {162},
number = {6},
pages = {1239-1249},
doi = {https://doi.org/10.1111/j.1476-5381.2010.01127.x},
year = {2011}
}

@article{dimasi2016innovation,
  title={{Innovation in the Pharmaceutical Industry: New Estimates of R\&D Costs}},
  author={DiMasi, Joseph A and Grabowski, Henry G and Hansen, Ronald W},
  journal={Journal of Health Economics},
  volume={47},
  pages={20--33},
  year={2016},
  publisher={Elsevier},
  doi={https://doi.org/10.1016/j.jhealeco.2016.01.012}
}

@article{schuhmacher2023analysis,
  title={{Analysis of Pharma R\&D Productivity--A New Perspective Needed}},
  author = {Alexander Schuhmacher and Markus Hinder and Alexander {von Stegmann und Stein} and Dominik Hartl and Oliver Gassmann},
  journal={Drug Discovery Today},
  volume={28},
  number={10},
  pages={103726},
  year={2023},
  publisher={Elsevier},
  doi = {10.1016/j.drudis.2023.103726}
}

@article{theodoris2023transfer,
  title={{Transfer Learning Enables Predictions in Network Biology}},
  author={Theodoris, Christina V and Xiao, Ling and Chopra, Anant and Chaffin, Mark D and Al Sayed, Zeina R and Hill, Matthew C and Mantineo, Helene and Brydon, Elizabeth M and Zeng, Zexian and Liu, X Shirley and others},
  journal={Nature},
  volume={618},
  number={7965},
  pages={616--624},
  year={2023},
  publisher={Nature Publishing Group UK London},
  doi = {10.1038/s41586-023-06139-9}
}

@article{wolf2018scanpy,
  title={{SCANPY: Large-Scale Single-Cell Gene Expression Data Analysis}},
  author={Wolf, F Alexander and Angerer, Philipp and Theis, Fabian J},
  journal={Genome Biology},
  volume={19},
  number={1},
  pages={15},
  year={2018},
  publisher={Springer},
  doi = {10.1186/s13059-017-1382-0}
}

@article{morgan2011cost,
  title={{The Cost of Drug Development: A Systematic Review}},
  author={Morgan, Steve and Grootendorst, Paul and Lexchin, Joel and Cunningham, Colleen and Greyson, Devon},
  journal={Health Policy},
  volume={100},
  number={1},
  pages={4--17},
  year={2011},
  publisher={Elsevier},
  doi = {https://doi.org/10.1016/j.healthpol.2010.12.002}
}

@article{frijters2010literature,
  title={{Literature Mining for the Discovery of Hidden Connections Between Drugs, Genes and Diseases}},
  author={Frijters, Raoul and Van Vugt, Marianne and Smeets, Ruben and Van Schaik, Ren{\'e} and De Vlieg, Jacob and Alkema, Wynand},
  journal={PLoS Computational Biology},
  volume={6},
  number={9},
  pages={e1000943},
  year={2010},
  publisher={Public Library of Science San Francisco, USA},
  doi = {10.1371/journal.pcbi.1000943}
}

@article{mayr2016deeptox,
  title={{DeepTox: Toxicity Prediction Using Deep Learning}},
  author={Mayr, Andreas and Klambauer, G{\"u}nter and Unterthiner, Thomas and Hochreiter, Sepp},
  journal={Frontiers in Environmental Science},
  volume={3},
  pages={80},
  year={2016},
  publisher={Frontiers},
  doi= {10.3389/fenvs.2015.00080}
}

@article{jimenez2018k,
  title={{KDEEP: Protein--Ligand Absolute Binding Affinity Prediction via 3D-Convolutional Neural Networks}},
  author={Jim{\'e}nez, Jos{\'e} and Skalic, Miha and Martinez-Rosell, Gerard and De Fabritiis, Gianni},
  journal={Journal of Chemical Information and Modeling},
  volume={58},
  number={2},
  pages={287--296},
  year={2018},
  publisher={ACS Publications},
  doi = {10.1021/acs.jcim.7b00650}
}

@article{elmarakeby2021biologically,
  title={{Biologically Informed Deep Neural Network for Prostate Cancer Discovery}},
  author={Elmarakeby, Haitham A and Hwang, Justin and Arafeh, Rand and Crowdis, Jett and Gang, Sydney and Liu, David and AlDubayan, Saud H and Salari, Keyan and Kregel, Steven and Richter, Camden and others},
  journal={Nature},
  volume={598},
  number={7880},
  pages={348--352},
  year={2021},
  publisher={Nature Publishing Group UK London},
  doi = {10.1038/s41586-021-03922-4}
}

@article{cui2024scgpt,
  title={{scGPT: Toward Building a Foundation Model for Single-Cell Multi-Omics Using Generative AI}},
  author={Cui, Haotian and Wang, Chloe and Maan, Hassaan and Pang, Kuan and Luo, Fengning and Duan, Nan and Wang, Bo},
  journal={Nature Methods},
  volume={21},
  number={8},
  pages={1470--1480},
  year={2024},
  publisher={Nature Publishing Group US New York},
  doi = {10.1038/s41592-024-02201-0}
}

@inproceedings{rozemberczki2022shapley,
  title={{The Shapley Value in Machine Learning}},
  author={Rozemberczki, Benedek and Watson, Lauren and Bayer, P{\'e}ter and Yang, Hao-Tsung and Kiss, Oliv{\'e}r and Nilsson, Sebastian and Sarkar, Rik},
  booktitle = {Proceedings of the Thirty-First International Joint Conference on Artificial Intelligence},
  pages={5572--5579},
  year={2022},
  doi = {10.24963/ijcai.2022/778},
  address={Vienna, Austria},
  publisher={International Joint Conferences on Artificial Intelligence Organization}
}

@article{mohs2017drug,
title = {{Drug Discovery and Development: Role of Basic Biological Research}},
journal = {Alzheimer's \& Dementia: Translational Research \& Clinical Interventions},
volume = {3},
number = {4},
pages = {651-657},
year = {2017},
issn = {2352-8737},
doi = {https://doi.org/10.1016/j.trci.2017.10.005},
author = {Richard C. Mohs and Nigel H. Greig},
}

@ARTICLE{wang2026TrajLens,
  author={Wang, Qipeng and Ruan, Shaolun and Sheng, Rui and Wang, Yong and Zhu, Min and Qu, Huamin},
  journal={IEEE Transactions on Visualization and Computer Graphics}, 
  title={{TrajLens: Visual Analysis for Constructing Cell Developmental Trajectories in Cross-Sample Exploration}}, 
  year={2026},
  volume={32},
  number={1},
  pages={1219-1229},
  doi={10.1109/TVCG.2025.3634875}}

@ARTICLE{sheng2025TrialCompass,
  author={Sheng, Rui and Wang, Xingbo and Wang, Jiachen and Jin, Xiaofu and Sheng, Zhonghua and Xu, Zhenxing and Rajendran, Suraj and Qu, Huamin and Wang, Fei},
  journal={IEEE Transactions on Visualization and Computer Graphics}, 
  title={{TrialCompass: Visual Analytics for Enhancing the Eligibility Criteria Design of Clinical Trials}}, 
  year={2026},
  volume={32},
  number={1},
  pages={1230-1240},
  doi={10.1109/TVCG.2025.3634803}}

\end{document}